\RequirePackage{fix-cm}
\documentclass[twocolumn]{svjour3}          
\smartqed  
\usepackage{graphicx}
\usepackage[colorlinks=true, linkcolor=blue, urlcolor=blue, citecolor=blue]{hyperref}
\usepackage{booktabs}
\usepackage{comment}
\usepackage{color, colortbl}
\definecolor{Gray1}{gray}{0.7}
\definecolor{Gray2}{gray}{0.9}
\usepackage{arydshln}
\usepackage{subfigure}
\usepackage{amsmath}
\usepackage{tcolorbox}
\DeclareUnicodeCharacter{00A0}{ }

\journalname{Preprint: Submitted to ArXiv}
\begin{document}

\title{CoVScreen: Pitfalls and recommendations for screening COVID-19 using Chest X-rays
}


\author{Sonit Singh
}

\institute{Sonit Singh \at
              School of Computer Science and Engineering \\
              UNSW Sydney, Australia \\
              \email{sonit.singh@unsw.edu.au}}


\maketitle

\begin{abstract}

The novel coronavirus (COVID-19), a highly infectious respiratory disease caused by the severe acute respiratory syndrome coronavirus 2 (SARS-CoV-2) has emerged as an unprecedented healthcare crisis. The pandemic had a devastating impact on the health, well-being, and economy of the global population. The COVID-19 brought immense pressure on healthcare systems of many countries and were struggling with scaling up capacities to meet rising number of infections. Early screening and diagnosis of symptomatic patients plays crucial role in isolation of patient to help stop community transmission as well as providing early treatment helping in reducing the mortality rate. Although, the real-time Reverse Transcription Polymerase Chain Reaction (RT-PCR) test is the gold standard for COVID-19 testing, it is a manual, laborious, time consuming, uncomfortable, and invasive process. Besides this, many countries faced the shortage of testing kits and in some cases, RT-PCR test has led to high false negatives. To overcome these challenges and to complement, clinicians recommends chest radiography as an alternative to RT-PCR testing. Due to its accessibility, availability, lower-cost, ease of sanitisation, and portable setup, chest X-Ray imaging can serve as an effective screening and diagnostic tool. In this study, we aim to augment radiologists by providing second opinion through computer-aided diagnosis and screening of chest X-rays using deep learning based image classification model. We first highlight limitations of existing datasets and studies in terms of data quality, data imbalance, and evaluation strategy. We curated a large-scale COVID-19 chest X-ray dataset from many publicly available COVID-19 imaging databases and proposed a pre-processing pipeline to improve quality of the dataset. We proposed  \emph{CoVScreen}, a Convolutional Neural Network (CNN) to train and test the curated dataset. The experimental results applying different classification scenarios on the curated dataset in terms of various evaluation metrics demonstrate the effectiveness of proposed methodology in the screening of COVID-19 infection. The proposed solution can effectively diagnose COVID-19 disease from chest radiographs, thereby providing potential benefits to radiology workflow and patient care in terms of screening, tracking the progression, and planning treatment of COVID-19 disease.

\keywords{Medical imaging \and chest X-ray \and medical image classification \and Coronavirus \and COVID-19 \and convolutional neural network \and computer-aided diagnosis}
\end{abstract}

\section{Introduction}\label{intro}

The novel coronavirus (COVID-19) is a respiratory infectious disease caused by severe acute respiratory syndrome coronavirus 2 (SARS-CoV-2)~\cite{Sharma:Tiwari:2020:sars-cov-2}. One of the key challenges for governments and health professionals is early screening and diagnosis of COVID-19 symptomatic patients. Being a highly contagious disease with fast community transmission, early screening of symptomatic patients can not only provide early treatment to the infected individual but also helps to avoid community transmission. Hence effective and early screening of COVID-19 infected patients is a critical and important step in fighting the pandemics such as of COVID-19. Apart from this, research studies shows that treatment of COVID-19 infected patients at an early stage helps in reducing the mortality rate~\cite{Sun:Qiu:2020:lower_mortality:}. 

The real-time Reverse Transcription Polymerase Chain Reaction (RT-PCR) is the gold standard for the COVID-19 testing. However, it is a manual, laborious, and time-consuming process. Also, it has issues having high false negative rate. Recent studies on COVID-19 using chest radiographs suggest that chest radiographs contain salient radiographic features specific to COVID-19~\cite{Punn:Agrawal:2020:automatic_diagnosis}. Given medical imaging has served as an effective screening aid for various thoracic diseases and since COVID-19 can lead to pneumonia and manifests in lungs, medical imaging in the form of chest X-rays can be mined for patterns that can provide such distinguishable factors for the identification of the virus. Researchers believe that chest radiology based system can be an effective tool in detection, quantification and follow-up of COVID-19 cases. Studies shows that the use of CT or X-rays to diagnose and screen COVID-19 can provide higher sensitivity and can be used as an alternative to RT-PCR test~\cite{Ng:Elaine:2020:imaging_profile}. 

In clinical practice, the RT-PCR test is usually complemented with a chest X-ray, in such as manner that the combined analysis reduces the significant number of false negatives, and at the same time, brings additional information about the extent and severity of the disease~\cite{Arias-Londono:2020:AI_applied_to_CXRs}. Hence, chest X-ray imaging has become the standard in the screening process since it is fast, minimally-invasive, low-cost, and require simpler logistics for its implementation. Medical imaging based COVID-19 diagnosis system can be fast, analyse multiple cases simultaneously, have greater availability and more importantly, such system can be very useful in hospitals with no or limited number of testing kits and resources.  Moreover, given the importance of radiography in modern health care system, radiology imaging systems are available in every hospital, thus making radiography based approach more convenient and easily available. If there are shortages of PCR test, it is important to turn to medical imaging to manage overwhelming patient load. In such context, AI based COVID-19 diagnosis represents timely risk stratification of patients. Therefore, there is a need for a computer-aided diagnosis system to help radiologists interpret images faster and accurately. 

Although less sensitive than Computed Tomography (CT) scans, Chest X-Rays (CXRs) are significantly more accessible and affordable in various hospital settings across the globe. Furthermore, the availability of portable CXR systems allow for imaging to be done within isolated rooms, significantly lowering the risk of spread of COVID-19 disease. The radiologists consider chest X-ray images over CT-scan as primary radiography examination to detect the infection caused by COVID-19 due to high availability of X-ray machines in most of the hospitals, low ionizing radiations, and low cost of X-ray machines compared to CT-scan machine. Apart from this, CT imaging takes more time than X-ray imaging, and real-time X-ray imaging can significantly accelerate the speed of disease screening. Multiple countries were using portable CXRs as a first-line triage tool for diagnosis as well as assessing the severity of the COVID-19 infection~\cite{Wong:Lam:2020:frequency_and_distribution}. 

While radiography in medical examinations can be quickly performed and become widely available with the prevalance of chest radiology imaging systems in healthcare systems, the interpretation of radiography images by radiologists is limited due to the human capacity in detecting the subtle visual features present in the images. Because AI can discover patterns in chest X-rays that normally would not be recognised by radiologists, it is important to develop computer-aided diagnosis and screening tools for COVID-19 screening and diagnosis. These CAD tools can assist clinicians in making optimal clinical decisions by saving time, resources, and improving productivity.  We did comprehensive literature review to identify studies in 2020 and 2021 when COVID-19 pandemic was at full rise and identified limitations. The purpose of this research is to highlight limitations of these studies in terms of datasets and experimental design, and recommended the need to have thoughtful evaluation to check the effectiveness of deep learning models in the automatic diagnosis of COVID-19 from chest X-Rays.

Artificial Intelligence (AI) techniques applied to chest radiographs have potential to accurately diagnose COVID-19 disease. It can also be assistive to overcome the problem of lack of specialised physicians in resources-scarce remote geographic locations. Although various studies reported high performance of deep learning models in diagnosing COVID-19 using chest X-rays, there are various limitations. First, most of the these studies worked on small number of COVID-19 chest X-rays. Since deep learning models are data hungry, these models can easily overfit on small number of samples~\cite{Yosinski:2014:how_transferable,romero:2020:targeted}. Second, majority of studies have imbalanced datasets and reported performance of their proposed methods in terms of accuracy score. Various research studies have shown that accuracy score is not a robust metric, specially on imbalanced datasets~\cite{He:2009:Learning_from_imbalanced_data,Krawczyk:2016:learning_from_imbalanced_data}. A model trained as a binary classifier having $25$ COVID-19 CXRs and $5000$ Normal CXRs can give high accuracy score, which is not the true performance of the trained model. Hence, it is important to work on approximately balanced dataset and also to report other classification metrics such as precision, recall, f1-score, and area under the ROC curve (AUC). Third, existing studies have focused on only three settings: 2-class classification (Normal vs. COVID-19); 3-class classification (Normal vs. Pneumonia vs. COVID-19); and 4-class classification (Normal vs. Bacterial Pneumonia vs. Viral Pneumonia vs. COVID-19). Since most of diseases do not occur in isolation, it is highly plausible that chest X-ray can have multiple medical conditions. Hence, existing studies ignore the co-occurrence of diseases in a single chest X-ray. Fourth, most of the studies doing 3-class classification and 4-class classification used pneumonia chest X-rays of paediatric patients of one to five years old. Since COVID-19 and other thoracic samples are of adult population but having pneumonia samples of children population, the trained model suffers from prediction bias problem, \textit{i.e.}, the model might make predictions based on chest size itself (given that children CXRs have smaller size than adult population).

To overcome the above stated limitations, we make the following contributions:
\begin{itemize}
    \item We proposed \emph{CoVScreen}, a deep learning model for automatic screening and diagnosis of COVID-19 infected patients. 
    \item We curated a relatively larger COVID-19 dataset from multiple publicly available databases.
    \item We did experiments to check model performance having chest X-rays from adult population and children to see whether model have bias towards learning dataset characteristics.
    \item We did experiments to check the effect of severity of disease on the diagnostic accuracy. In this study, we took samples of different severity labels, making it robust to check model performance for different severity levels of the disease. 
\end{itemize}

\section{Related Work}\label{related_work_section}
There are several published studies on chest X-rays based COVID-19 diagnosis. Here, we briefly review representative studies focused on COVID-19 diagnosis in the year 2020 and 2021. The  success  of  AI  based  techniques  in automatic  diagnosis  in  the  medical  field  and rapid rise in COVID-19 cases have necessitated the need of AI based automatic detection and diagnosis system. Recently, many researchers have used radiology images for COVID-19 detection. \cite{wang:2020:covidnet} introduced COVID-Net, a deep CNN designed for the detection of patients with COVID-19 from chest X-ray images that is publicly available. COVID-Net achieved an overall accuracy of 83.5\% for 4-classes (COVID-19, Bacterial pneumonia, Viral pneumonia, normal) and 92.4\% in classifying COVID-19, normal and non-COVID pneumonia cases. \cite{narin:2020:automatic} evaluated different CNN architectures for the diagnosis of COVID-19 and achieved an accuracy of 98\% using a pre-trained ResNet-50 model. However, the classification problem is overly simplified by only discriminating between healthy and COVID-19 patients, disregarding the problem of discriminating regular pneumonia conditions from COVID-19 conditions. \cite{wang:2020:covidnet} proposed COVID-Net to detect distinctive abnormalities in CXR images of COVID-19 patients among samples of patients with non-COVID-19 viral infections, bacterial infections, and healthy patients.  \cite{ghoshal:2020:estimating} employed uncertainty estimation and interpretability based on Bayesian approach to CXR based COVID-19 diagnosis. \cite{hemdan:2020:covidxnet} introduced COVIDX-Net to aid radiologists to automatically detect COVID-19 in chest X-rays. The model is validated on 50 chest X-rays comprising 25 cases of COVID-19 and 25 cases without any infections. This study reveals that The VGG19 and DenseNet models have similar performance of automated COVID-19 detection with f1-scores of 0.91 and 0.89 for COVID-19 and normal, respectively. \cite{sethy:2020:covid-19} proposed CNN based system for detecting COVID-19 from chest X-rays. This model is trained on a dataset collected from Github, Kaggle and Open-I repositories and achieved an accuracy of 95.38\% for detecting COVID-19. \cite{ozturk:2020:DarkNet} proposed DarkCovidNet for automatic diagnosis of multi-class classification (COVID vs. Normal vs. Pneumonia) and binary classification (COVID vs. Normal). The DarkCovidNEt achieved accuracy score of $87.02\%$ for multi-class classification and $98.08\%$ for binary classification.
\cite{apostolopoulos:2020} trained different pre-trained CNN models on two different datasets. First dataset have 1427 chest X-rays (504 normal, 700 bacterial pneumonia, and 224 COVID-19). The proposed model achieved an accuracy of $98.75\%$ and $93.48\%$ for binary and 3-class classification. 
\cite{khan:2020:CoroNet} proposed CoroNet, a CNN to detect COVID-19 using Chest X-rays and CT scans. This model is based on Xception architecture and pre-trained on ImageNet dataset. The experimental results provides an accuracy of $89.6\%$ for 4-class classification (pneumonia viral vs. COVID-19 vs. pneumonia bacterial vs. normal) and an accuracy of $95\%$ for 3-class classification (normal vs. COVID-19 vs. pneumonia). \cite{xu:2020:screening-coronavirus} introduced a screening tool to differentiate COVID-19 from viral pneumonia and normal cases using $618$ pulmonary CT samples (175 normal; 224 viral pneumonia; and 219 COVID-19 patients). This model achieved a total accuracy of $86.7\%$.
\cite{wang:2020:covid19-CT} proposed a deep CNN model to classify COVID-19 from viral pneumonia using $99$ chest CT samples (55 viral pneumonia and 44 COVID-19). The results of testing set show an overall accuracy of $73.1\%$, along with a sensitivity of $74.0\%$ and a specificity of $67.0\%$. \cite{li:2020:CT} trained a ResNet-50 model to distinguish COVID-19 from pneumonia and non-pneumonia using 4356 chest CT images (1735 pneumonia, 1325 non-pneumonia and 1296 COVID-19). The results show that the COVNet model provides a specificity of $96\%$, a sensitivity of $90\%$, and AUC of $0.96$ in classifying COVID-19.
\cite{song:2020} introduced DeepPneumonia, a system based on deep learning to identify patients with COVID-19 from healthy and bacterial pneumonia patients using chest CT images. This model achieves an overall accuracy of $86.0\%$ for (COVID-19 vs. bacterial pneumonia) classification and an overall accuracy of $94.0\%$ for (COVID-19 vs. healthy) classification. \cite{ghoshal:2020:estimating} trained a Bayesian deep learning classifier using transfer learning method to estimate model uncertainty using COVID-19 chest X-ray images. \cite{khalifa:2020:detection} used generative adversarial networks (GANs) to address the overfitting problem for detection pneumonia from CXR images. \cite{Bukhari:2020:the_diagnostic_evaluation} used ResNet-50 model for the diagnosis of COVID-19 infection on total 278 CXRs, having normal, pneumonia, and COVID-19 cases. Although the study demonstrate promising results, the study has limitations in terms of limited number samples. 
\cite{farooq:2020:covidresnet} proposed an improved ResNet model, named \emph{COVIDResNet} for the screening of COVID-19 infection. They used different training techniques including progressive resizing, cycling learning rate, and discriminative learning rates to gain fast and accurate training. 
\cite{Zhang:Xie:2020:viral_pneumonia_anomaly_detection} developed an anomaly detection based algorithm for distinguishing COVID-19 from other types of pneumonia. All of the studies in literature advocate the need of automated system for the screening and diagnosis of COVID-19 infected patients. Although some of these studies showed promising results, they have not addressed the variable quality of the COVID-19 samples to minimise sampling bias. We also noted that datasets used in these studies are quite small and unbalanced. Class imbalance is a common problem in medical imaging datasets that can lead to inflated results because of over-classification of majority class at the expense of under-classification of the minority class. 

\subsection{Limitations of existing work in the literature and our contributions}
Despite the excellent results reported by several studies in the literature, our analysis reveals that there are certain limitations in the existing studies, in terms of their database curation, experimental design, evaluation criteria, and conclusions drawn. These limitations make it hard to translate existing research in the actual clinical settings. 

First, most of the existing studies worked on small number of images. Since deep learning models can easily overfit on small datasets, it does not provide clear image of model's generalisation capabilities. In general, most of the studies work on less than 200 chest X-rays having COVID-19 infection. However, this is understandable given that some of these studies were undertaken during the onset of COVID-19 pandemic when chest X-rays for COVID-19 were hardly available or numbers were quite low. To get a better understanding of model's generalisation capability, there is a need to have study involving large number of COVID-19 chest X-rays. 

Second, most of the studies took \emph{normal} chest X-rays from the Paul Mooney pneumonia challenge hosted at Kaggle~\cite{kaggle:paul_mooney_chest_X-ray}. Also, for the 3-class classification (Normal vs. Pneumonia vs. COVID-19) and 4-class classification (Normal vs. Bacterial Pneumonia vs. Viral Pneumonia vs. COVID-19), most of these studies took chest X-rays from the pneumonia challenge dataset. The pneumonia detection challenge contains chest X-rays for paediatric population having age of 1 to 5, collected at Guangzhou hospital, China. Since, machine learning model can learn anatomy-specific features easily given a mix of population, it is possible that the reported results of most of the existing studies bias. Therefore, it is important to check whether this is the case for the existing studies and steps will be taken to overcome this data specific bias in machine learning models. 

Third, most of the existing studies worked on \emph{imbalanced} dataset having small number of COVID-19 chest X-rays whereas quite large number of chest X-rays for normal or pneumonia class. Reporting the accuracy score on an imbalanced dataset is not a robust metric and do not indicate model's true performance. Given most of the studies in the literature reported results on imbalanced dataset with over 95\% accuracy, it is hard to comprehend whether high accuracy is just because of majority class without model knowing anything about the minority class. To have a robust measure of the model performance, it is important to work on approximately balanced data and reporting results in terms of different metrics, including precision, recall, f1-score, and AUC score. 

Fourth, most of the existing studies worked on two-class (Normal vs. COVID-19), three-class (Normal vs. Pneumonia vs. COVID-19), and four-class (Normal vs. Bacterial pneumonia vs. Viral pneumonia vs. COVID-19). However, no existing studies consider doing classification of multiple thoracic diseases, which are highly likely to co-occur. Given most of the thoracic diseases do not occur in isolation, it is plausible to have the occurrence of multiple thoracic medical conditions which can be captured using chest X-rays. Hence, it is important to carry out analysis based on different thoracic diseases along with COVID-19 to check how well model can differentiate COVID-19 from other thoracic diseases. 

Fifth, most of the existing studies do not consider the severity of COVID-19 infection in chest X-rays. Most of the COVID-19 chest X-rays are coming from a combination of sources, repositories, different geographic locations, and having chest X-ray captured at different times from the onset of symptoms, which certainly affect the results. We hypothesise that model trained for binary classification (normal vs. COVID-19) performance will be low for chest X-rays captured when patients have not fully developed COVID-19 symptoms (mild symptoms). On the other hand, model performance is plausibly high for chest X-rays when patients have COVID-19 infection at its peak (severe disease symptoms). To validate this hypothesis, there is a need to have thorough study where chest X-rays have varying COVID-19 symptoms, from mild to severe case. 

Sixth, most of the existing studies have reported results in terms of accuracy scores, but have not provide any qualitative analysis of model's performance in terms of interpretability and explainability. Hence, it is unclear whether the good results are due to the system's actual capability to extract information related to the pathology or because it learnt other data specific features during training that are biasing and compromising the results. There are certain non-pathological features that are relevant for the classification task but might lack diagnostic relevance. Therefore, it is important to make use of different explainable and visualisation tools to provide explanation for model decisions. This also helps to establish health professionals' appropriate trust in model's capabilities in diagnosing certain medical conditions, including COVID-19 disease. 

In consideration to the above highlighted limitations of the existing studies as per the literature, it is important to carry out a study that addresses these concerns. In this paper, we aim to overcome some of the stated limitations so that model's true performance in diagnosing and screening the COVID-19 disease using chest X-rays can be reported. 

\section{Material and Methods}
In this section, we provide description about how we collected the dataset(s) and the COVScreen model architecture. 

\subsection{Dataset development}\label{dataset_development_section}
Since COVID-19 is a new disease, there is no appropriate sized dataset available that can be used for this study. In this research, we construct a representative large-scale dataset by collecting chest X-rays from multiple publicly available chest X-ray databases, which is a much larger and approximately balanced than previous studies. This section provides sources of data, data cleaning, and formulation of data for different experimental settings.

\subsubsection{Data Sources}
\begin{enumerate}
    \item IEEE8023/COVID-ChestXray-Dataset~\cite{cohen:2020:covid}: The Covid-ChestXray-dataset is a public database of pneumonia cases with CXRs related to COVID-19, MERS, SARS, and ARDS collected by Paul Cohen from multiple sites across the world. As of 20 January 2021, the dataset includes more than $290$ chest X-rays and CT scans of patients infected with COVID-19 and other diseases including SARS, MERS, and ARDS. In this study, we only focused on CXRs of patients having COVID-19 disease. The images in this dataset have discrepancies in terms of variable size (varying from $255 \times 249$ to $4280 \times 3520$) and quality in terms of brightness and subject positioning. The database is regularly updated with images shared by researchers from different regions across the globe. 
    
    \item Figure 1 COVID-19 Chest X-Ray Dataset Initiative~\cite{figure1_COVID-19_Dataset_Initiative}: This dataset has been publicly released recently and primarily contains COVID-19 cases. The dataset is curated by a team of researchers from DarwinAI corp, Canada, University of Waterloo, Canada, and Figure 1. 
    
    \item COVID-19 chest imaging cases in a hospital in Spain~\cite{twitter_spain_radiologists_cases}: This dataset contains CXRs collected from a hospital in Spain and primarily contains COVID-19 positive cases diagnosed with the RT-PCR test. The images are publicly shared by one of the cardiothoracic radiologist by Twitter. 
    
    \item SIRM COVID-19 Database~\cite{sirm_Italy_radiologists_cases}: This database contain CXRs of COVID-19 patients confirmed by RT-PCR test in Italy and are publicly shared by the Italian Society of Medical and Interventional Radiology (SIRM). We only included cases where chest X-rays are available for the study and case is confirmed positive by the swab test. We do not included cases of pediatric patients and cases for which there are no chest X-rays available. 
    
    \item Radiopedia~\cite{radiopaedia:COVID-19-cases}: The radiopaedia COVID-19 database consists of chest X-rays and CT scans of COVID-19 cases confirmed by the RT-PCR test. For this study, we only consider chest X-rays.
    
    \item COVIDGR-1.0 Dataset~\cite{Tabik:2020:COVIDGR}: The COVIDGR-1.0 dataset consists of chest X-rays to assist the diagnosis of COVID-19 and is built in close collaboration with radiology experts in Spain. All COVID-19 CXRs correspond to patients who have COVID-19 confirmed with RT-PCR test with less than 24 hours difference between acquisition of CXR and performing the RT-PCR test. The first version of dataset consists of total 852 CXRS (COVID-19: 426 and Normal:426). The COVID-19 cases cover the entire spectrum of severity scores having 76 normal cases, 100 mildly severe cases, 171 moderate cases, and 79 severe cases. Thus, COVIDGR dataset allows us to build more robust models that have ability to distinguish CXRs with different progression of disease. 
    
    \item Hannover Medical School COVID-19 Database~\cite{Winther:2020:Figshare-COVID-19-image-repository}: This COVID-19 images database consist of anonymised chest X-Rays of COVID-19 cases from the Institute for Diagnostic and Interventional Radiology, Hannover Medical School, Germany. 
    
    \item European Society of Radiology (ESR) COVID-19 cases: The European Society of Radiology~\footnote{\url{https://www.eurorad.org/advanced-search?search=COVID}} provides a learning environment for radiologists, radiology residents, and students worldwide. It is a peer-reviewed educational tool based on radiological case including COVID-19. 
    
    \item Peer-reviewed publications: Many peer-reviewed publications have discussed COVID-19 cases in terms of their radiographic findings and progression of disease for treatment planning~\cite{Jacobi:2020:portable_CXR_COVID19,Qian:2020:severe_acute_respiratory}. We collected these chest X-Rays from the COVID-CXNet~\cite{haghanifar:2020:COVID-CXNet} Github repository~\footnote{https://github.com/armiro/COVID-CXNet} by including only those images that appeared in peer-reviewed publications only. 
    
    \item BIMCV COVID19 Database~\cite{vaya:2020:bimcv}: The BIMCV COVID19 database consists of chest radiographs of COVID-19 patients along with results of PCR, collected by the Valencia Region Medical Image Bank (BIMCV), Spain. The database provides chest X-rays and CT scans, along with at least one positive PCR test, and radiological reports. 
    
    \item COVID-19 Chest X-ray image repository~\cite{haghanifar:2020:COVID-CXNet}: The COVID-19 Chest X-ray image repository have images from COVID-19 positive patients, confirmed by RT-PCR positive. 
    
\end{enumerate}

From the above COVID-19 radiographic databases, we curated a total of $4454$ chest X-rays for COVID-19. Table~\ref{tab:count_COVID19} provides count of COVID-19 CXRs from various databases.

\begin{table}
    \centering
    \caption{Count of COVID-19 CXRs from different COVID-19 radiographic databases.}
    \label{tab:count_COVID19}
    \begin{tabular}{rlr}
    \toprule
      \textbf{S.No.} & \textbf{Data source}  & \textbf{\#Images} \\ \midrule
    1 & IEEE8023 COVID-ChestX-Ray & 196 \\
    2 & Figure 1 & 35 \\
    3 & ActualMed & 58 \\
    4 & Twitter & 132 \\
    5 & SIRM & 38 \\
    6 & Radiopedia & 28 \\
    7 & COVIDGR1.0 & 426 \\ 
    8 & Hannover medical school & 243 \\
    9 & Eurorad & 258\\
    10 & Peer-reviewed publications & 166 \\ 
    11 & BIMCV & 2474 \\
    12 & COVID-19 CXR image repository & 400 \\\midrule
    & Total & 4454 \\ \bottomrule
    \end{tabular}
\end{table}
    
Th following sources are considered for collecting normal (healthy) cases as well as other thoracic disease labels except COVID-19. 
\begin{enumerate}
     \item NLM Montgomery County CXR Database~\cite{Jaeger:2014:two_public_chest}: This dataset consist of 138 posterior-anterior chest X-rays samples of tuberculosis and normal cases. The National Library of Medicine created this dataset in collaboration with the Department of Health and Human Services, Montgomery County, Maryland, USA. It contains data from chest X-rays collected under Montgomery County's tuberculosis screening program. 
    
    \item NLM Shenzhen (China) CXR Database~\cite{Jaeger:2014:two_public_chest}: This chest X-ray tuberculosis database is created by the National Library of Medicine, USA, in collaboration with Guangdong Medical College, Shenzhen, China. This database consists of 336 cases with manifestation of tuberculosis and 326 normal (healthy) cases. 
    
     \item RSNA Pneumonia Detection challenge Dataset~\cite{rsna_pneumonia_challenge}: The Radiological Society of North America (RSNA) Pneumonia detection challenge is a competition run by Kaggle organisers to build predictive models to distinguish pneumonia cases from normal cases. The RSNA challenge dataset is a subset of NIH ChestX-Ray14 dataset. Stage I of competition contains 25,684 training images and 1,000 testing images. For stage 2 of competition, organisers add 1,000 testing images from stage 1 into the training set and added 3,000 new testing images. Hence, the stage 2 dataset consists of 26,684 training images. Since the competition organisers provided annotations for only the training set, we use training images in our experimental settings. The CXRs are provided in DICOM format (.dcm files) and are converted to PNG format in order to have consistency with COVID-19 CXRs and to reduce computational cost. 
    
    \item ChestX-Ray14~\cite{Wang:2017:ChestX-ray14}: The National Institute of Health (NIH) ChestX-Ray14 dataset comprised of 112,120 chest X-Rays with $14$ common thoracic radiological findings. The dataset also contains normal cases ``No Findings" without specific findings in their corresponding images.  Although there exist other large-scale publicly available CXR databases including CheXpert~\cite{Irvin:2019:CheXpert}, MIMIC-CXR~\cite{Johnson:2019:MIMIC-CXR}, and PadChest~\cite{Bustos:2020:PadChest}; the ground truth labels of these datasets contains uncertain labels as they are automatically extracted from radiology report using NLP tools. The ChestX-Ray14 dataset provides images of uniform image size and labels do not contain any uncertain labels. For this study, studies with label ``No Findings" are used to be part of \emph{normal} class.
\end{enumerate}

\subsection{Data pre-processing}
\begin{figure*}
  \setcounter{subfigure}{0}
  \centering
  \subfigure[Side markers]{\includegraphics[width=3.5cm, height=3cm]{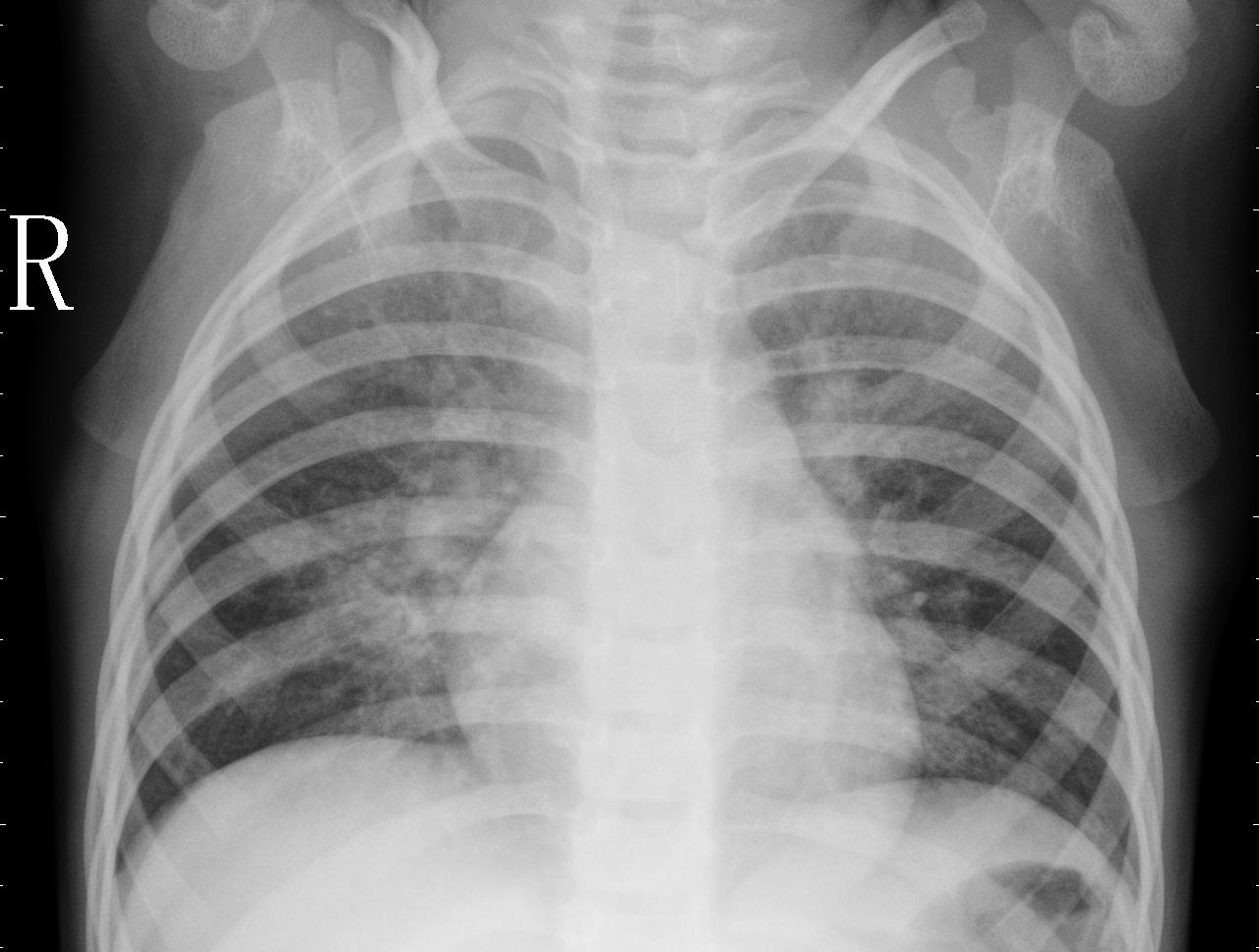}}\quad
  \subfigure[Low resolution]{\includegraphics[width=3.5cm, height=3cm]{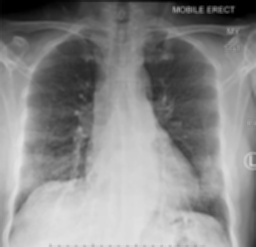}}\quad
  \subfigure[Children X-ray]{\includegraphics[width=3.5cm, height=3cm]{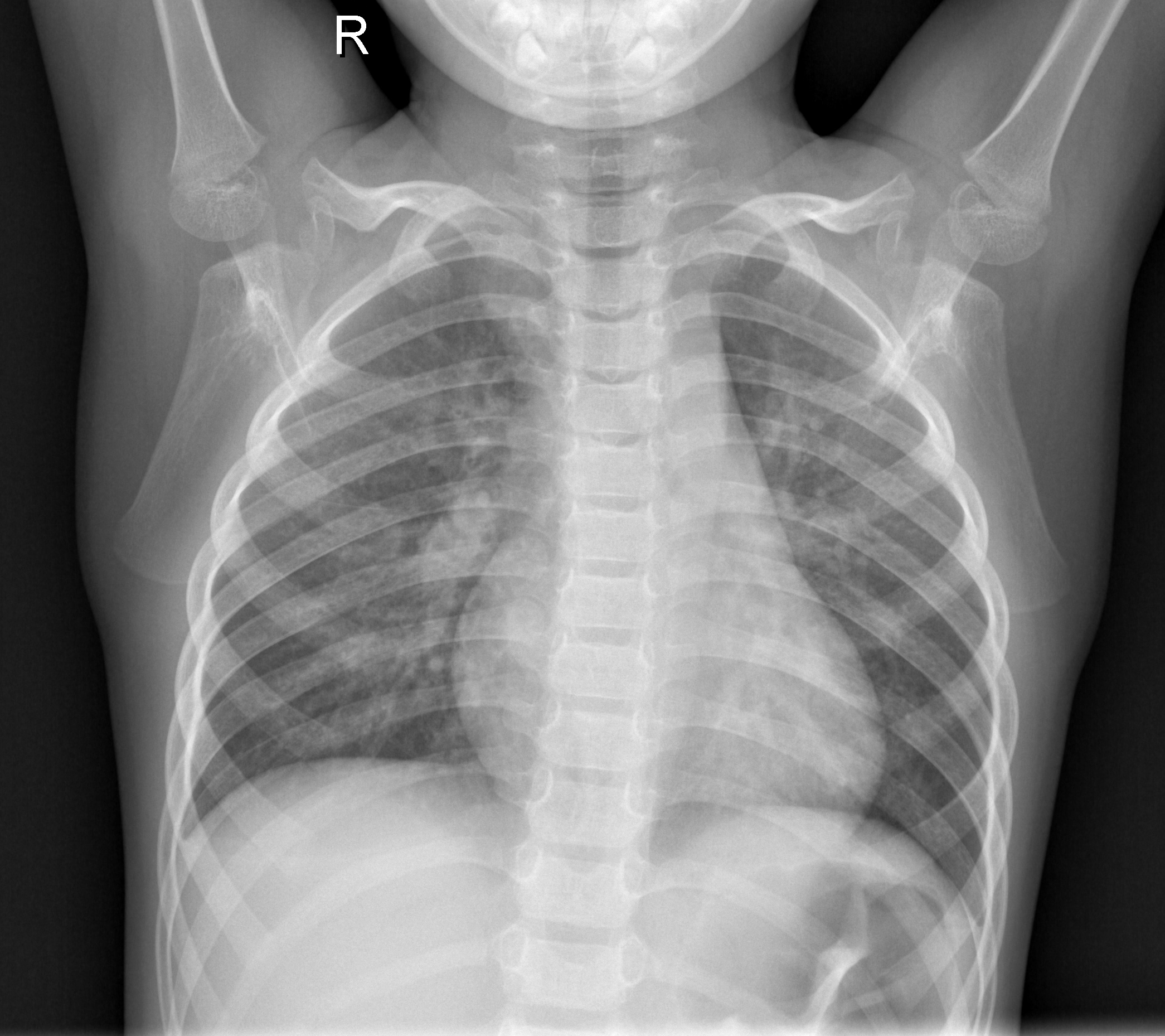}}\quad
  \subfigure[Low contrast]{\includegraphics[width=3.5cm, height=3cm]{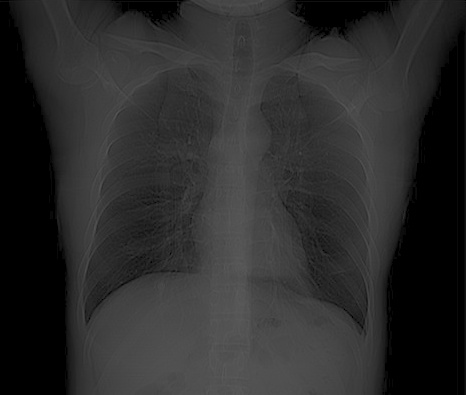}}\quad \\
  \subfigure[High contrast]{\includegraphics[width=3.5cm, height=3cm]{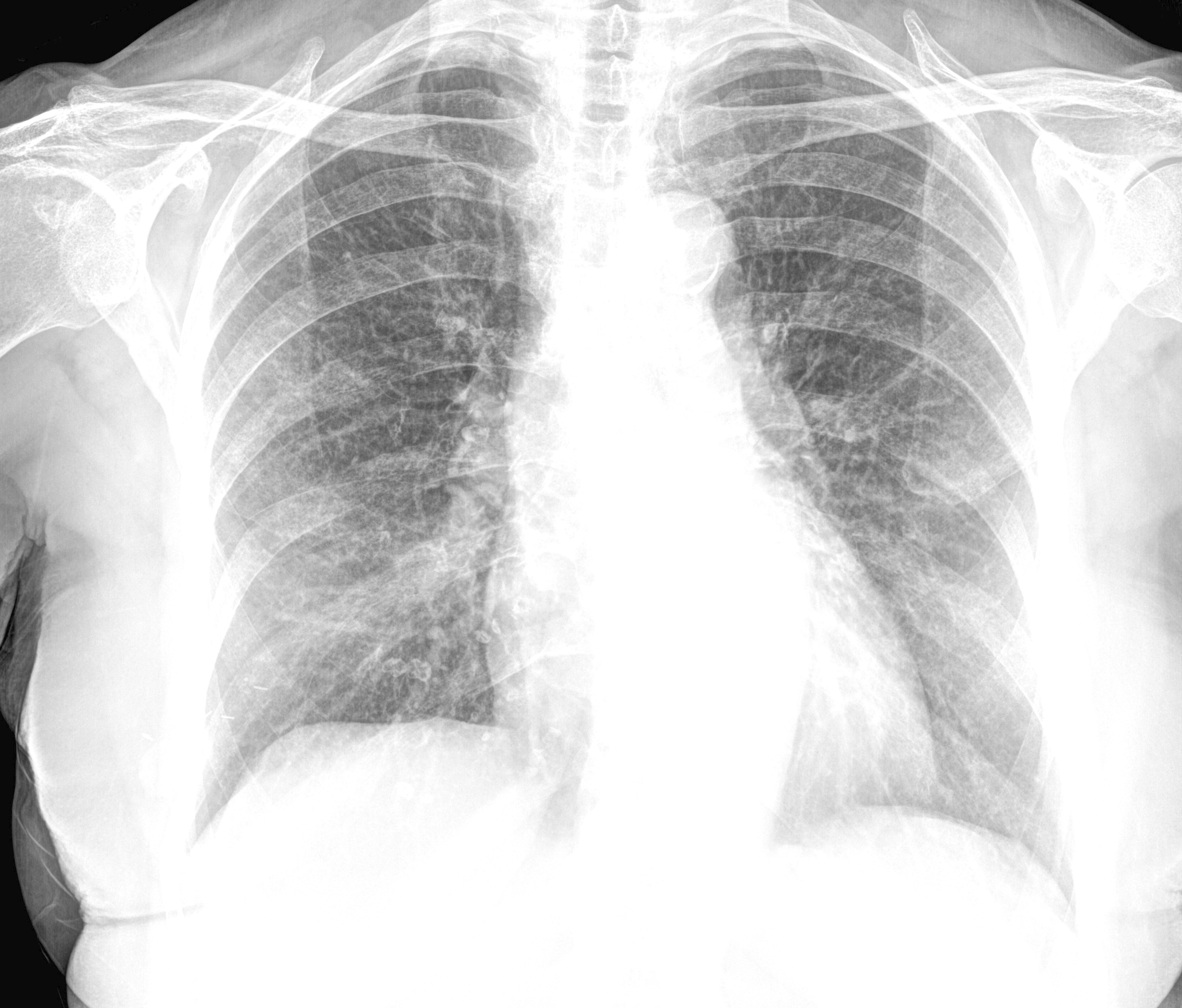}}\quad
  \subfigure[Squared-off]{\includegraphics[width=3.5cm, height=3cm]{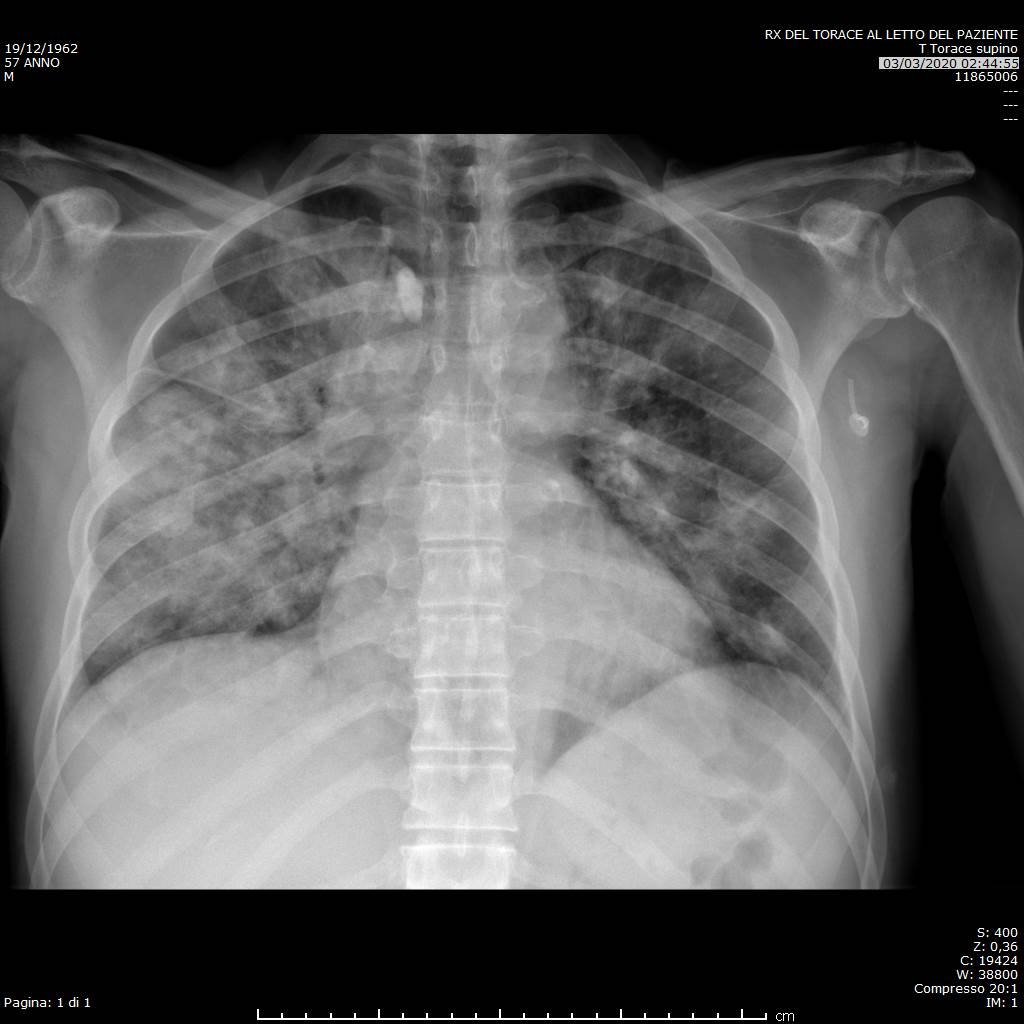}} \quad
  \subfigure[Noisy image]{\includegraphics[width=3.5cm, height=3cm]{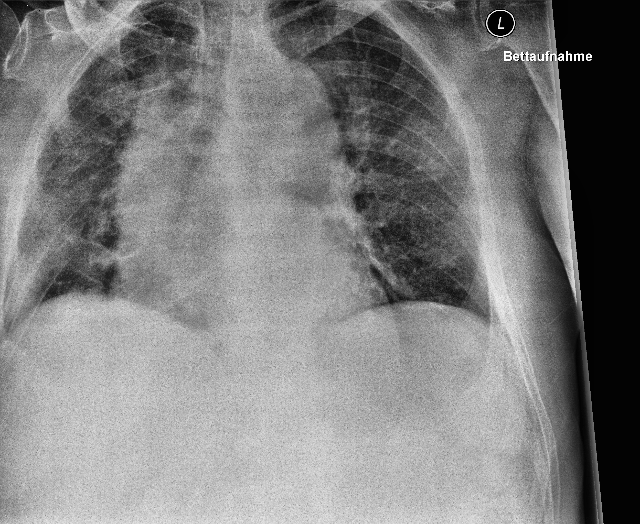}} \quad
  \caption{Diversity and irregularities in CXRs of COVID-19 dataset.} 
  \label{fig:diversity_images}
\end{figure*}

The quality of chest X-rays in the COVID-19 dataset is highly inconsistent since these images are shared from different parts of the world and are often taken in non-ideal conditions and vastly different equipment~\cite{horry:chakraborty:2020:X-ray}. On the other hand, ChestX-ray14 dataset showed uniformity and appeared to be highly curated. The discrepancy between these datasets are mostly ignored by current available studies. This may have impacted the performance of the models they developed, since there is a good chance that deep learning models learned the quality discrepancies between datasets rather than the characteristics of diseases. Figure~\ref{fig:diversity_images} shows various data irregularities which may give bias results. Having reviewed the related work in Section~\ref{related_work_section}, it is evident that despite the success of deep learning models in the detection of COVID-19 from chest radiographs, data irregularities have not been addressed. To reduce the impact of sampling bias, we implemented data pre-processing pipeline. The proposed pre-processing stage helps in improving the quality of visual information of each input image by reducing noise, improving contrast, removing extreme low or high frequencies, as well as removing textual side markers. 

\begin{figure}
\setcounter{subfigure}{0}
\centering
\subfigure[Original image]{
\includegraphics[width=.225\textwidth]{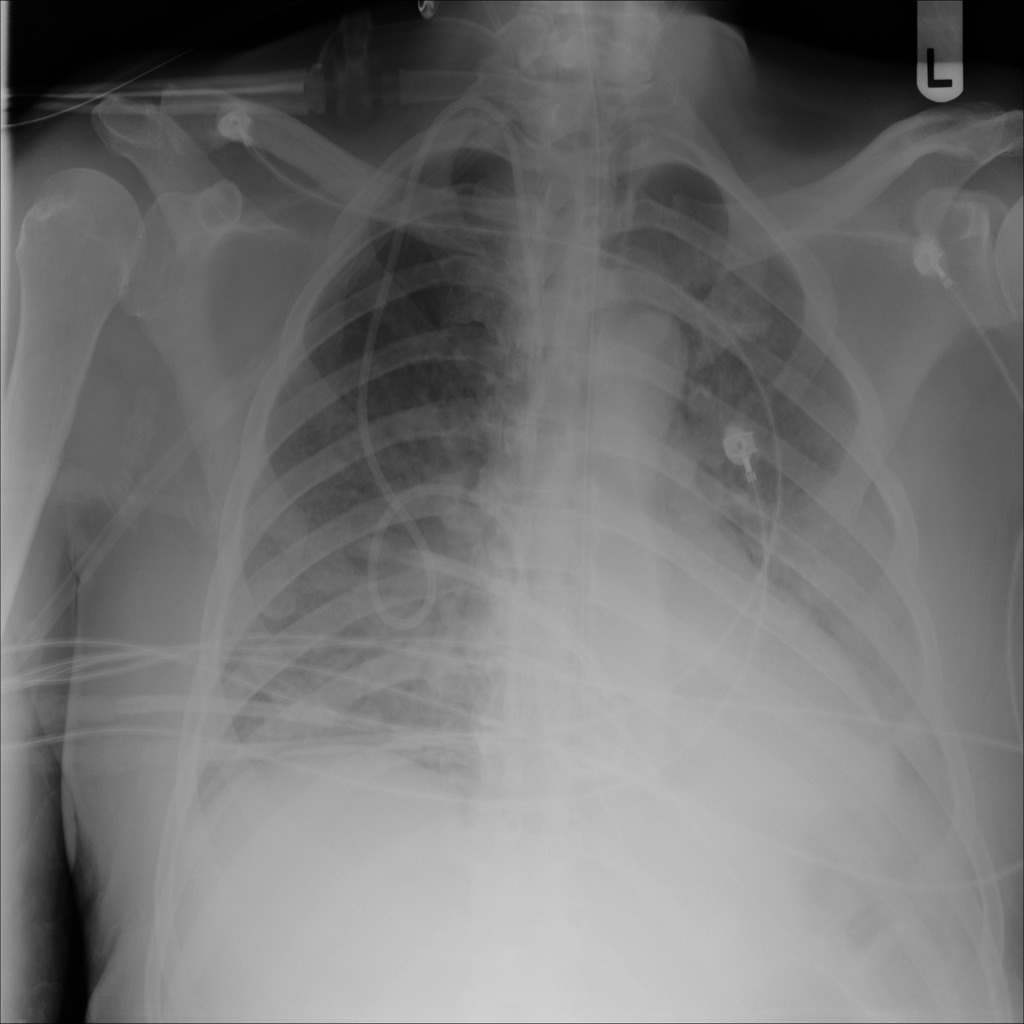}
}
\subfigure[Histogram]{
\includegraphics[width=.225\textwidth, height=3.8cm]{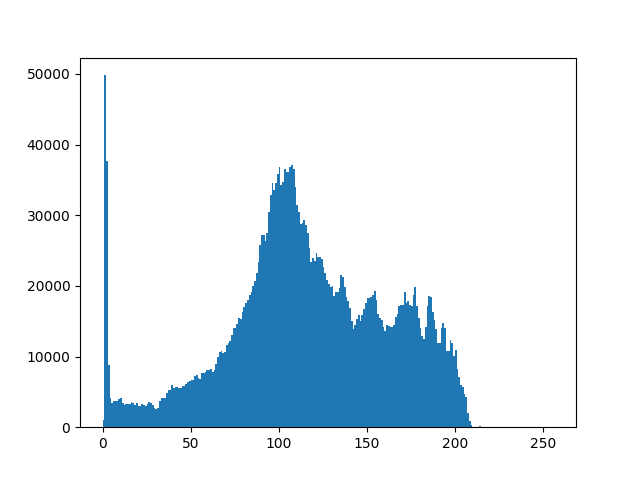}
}
\subfigure[Normalised image]{
\includegraphics[width=.225\textwidth]{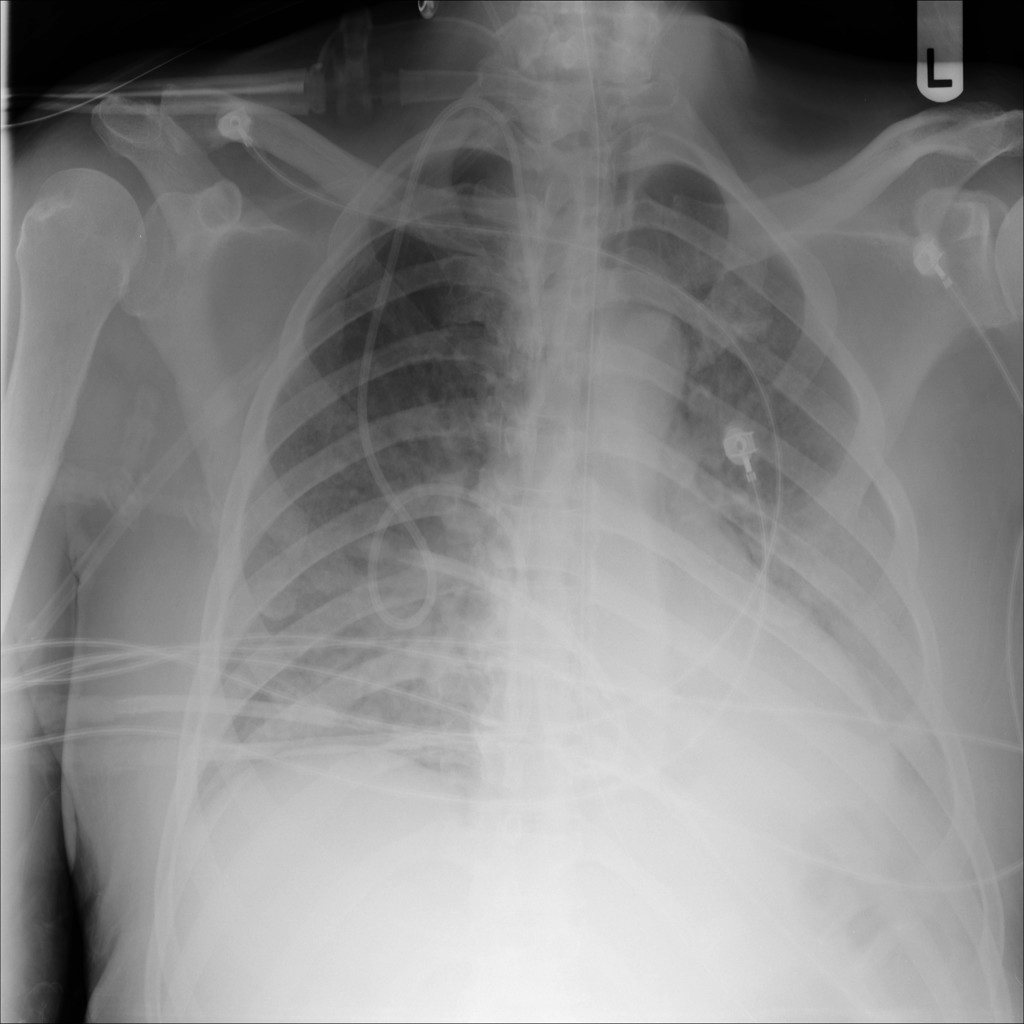}
}
\subfigure[CLAHE final output]{
\includegraphics[width=.225\textwidth]{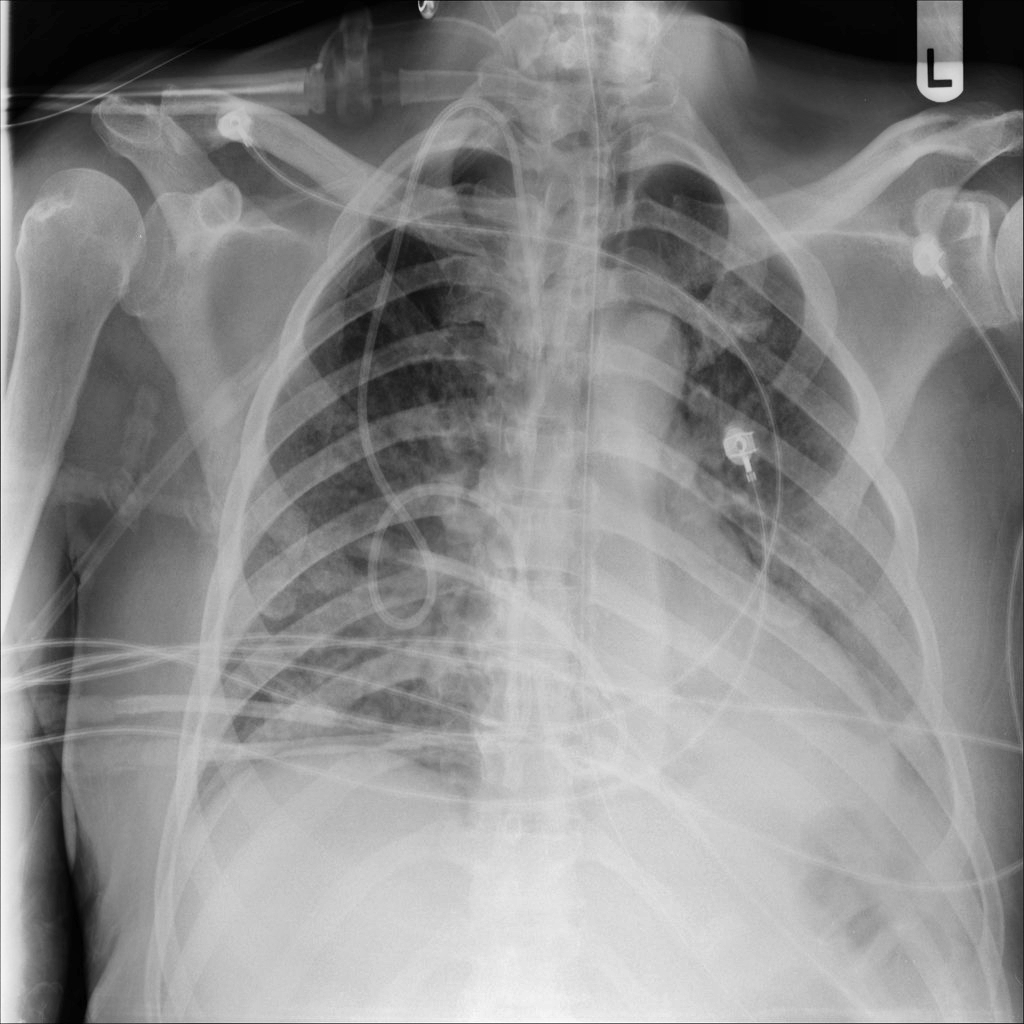}
}
\caption{Results of data pre-processing pipeline.}
\label{fig:preprocessing}
\end{figure}

\begin{enumerate}
    \item Intensity normalisation: We used OpenCV~\cite{opencv_library} library to normalise input images to the standard range between $0$ and $1$. 
    \item Histogram equalisation: Research shows that machine learning models work better when the intensity of tissue remains same for all images and not vary with the location of the tissue~\cite{Song:Zheng:2017:review_bias_correction}. Since CNNs use a pixel array as a data source, any systematic differences in the pixel intensity between datasets would introduce sampling bias in the results. This would have the consequence of training CNN on systematic image histogram differences rather than clinical disease features~\cite{horry:chakraborty:2020:X-ray}. In order to improve contrast of images and to delete low or high frequencies in images, we applied Contrast Limited Adaptive Histogram Equalisation (CLAHE)~\cite{Kerel:1994:CLAHE} technique. Figure~\ref{fig:preprocessing} shows improved image quality after applying intensity normalisation and contrast enhancement.
    \item Removing markers: The chest X-rays contains various position or orientation markers which can lead to bias as CNNs can fit to these extraneous features. In order to remove these text markers, we applied the following pipeline.
    \begin{enumerate}
        \item Obtain binary image: After loading the image, we applied low-pass filter and high-pass filter to remove any noise or high frequency content. Then, we applied \emph{Otsu} based threshold~\cite{Otsu:1979_threshold} to get a binary image.
        \item Getting text contours: Various position markers including \texttt{AP}, \texttt{PORTABLE}, and \texttt{SUPINE} are group of characters. In order to extract text as a single contour, we applied different \emph{morphological operations} to connect individual text contours to form a single contour.
        \item Text extraction: The previous step can return many contours. We filter contours based on their area and aspect ratio since markers have certain aspect ratio. We then draw bounding boxes over potential text markers. After this we perform bitwise operations to extract position markers.
        \item Image inpainting: In order to restore original image, we used \emph{inpainting} procedure using a Fast Marching Method (FMM)~\cite{Telea:2004:image_inpainting_FMM}. The FMM starts from the boundary of region to be restored and goes inside the region gradually filling everything in the boundary first. It takes a small neighbourhood around the pixel on the neighbourhood to be inpainted. This pixel is replaced by normalised weighted sum of all the known pixels in the neighbourhood.
    \end{enumerate}
\end{enumerate}

\begin{figure}
\setcounter{subfigure}{0}
\centering
\subfigure[Original image with marker]{
\includegraphics[width=.225\textwidth]{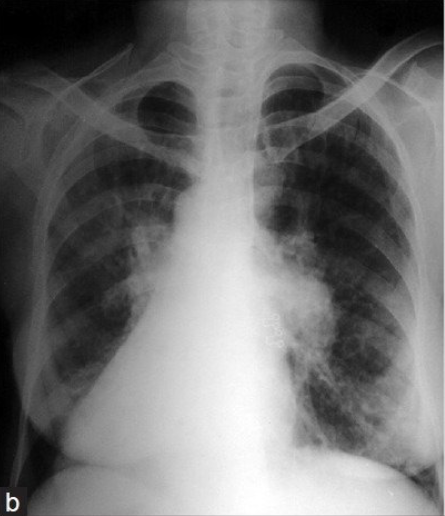}
}
\subfigure[Detected marker]{
\includegraphics[width=.225\textwidth]{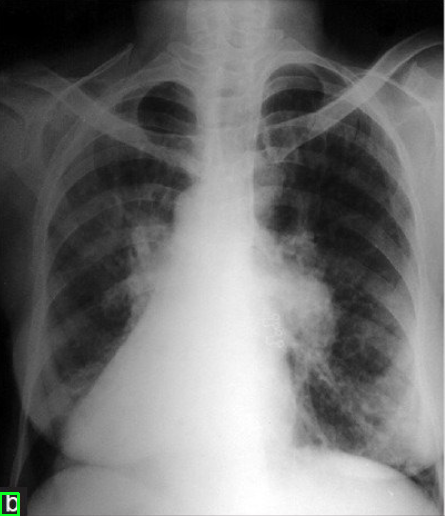}
}
\subfigure[Extracted marker]{
\includegraphics[width=.225\textwidth]{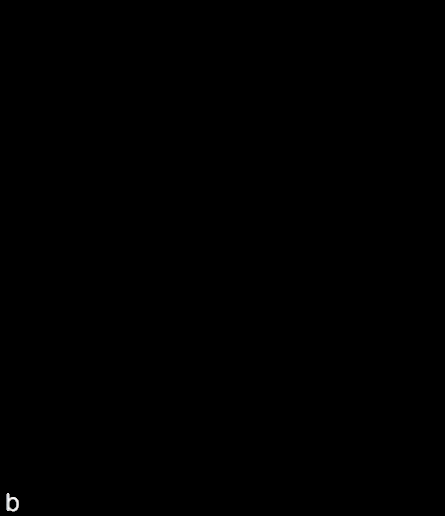}
}
\subfigure[Output image without marker]{
\includegraphics[width=.225\textwidth]{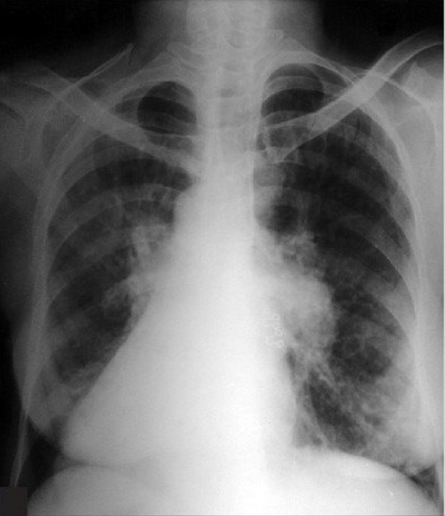}
}
\caption{Results of markers removal pipeline.}
\label{fig:markers_removal}
\end{figure}

All of the above points form our data pre-processing pipeline, which results in improved data quality of the curated dataset. Figure~\ref{fig:markers_removal} highlights intermediate and final results of markers removal pipeline.

\subsection{Data subset selection}
After data curation and pre-processing, we prepare subsets of data for different classification schemes. We train the proposed CoVScreen model for different classification schemes to check it effectiveness in discriminating COVID-19 infection from other thoracic diseases. Since medical imaging datasets have unequal number of samples for different classes, we applied different techniques to avoid data imbalance problem. Where it is possible to have approximately balanced dataset, we try to have equal number of sample for different classes. In certain cases, where it is not possible to have nearly perfect balanced dataset, we use class weightage scheme to alter weights of each training sample for a certain class depending upon the number of samples for a particular class within the entire dataset. 

The dataset subset contains $4454$ chest X-rays of COVID-19 infection and $4454$ chest X-rays of healthy subjects. This subset is designed for binary classification. The COVID-19 CXRs are taken from the complete COVID-19 whereas healthy cases are taken from ChestX-Ray14 ``No Findings" class. Table~\ref{tab:2-class_split} provides number of samples for training, validation, and testing in this dataset. 

\begin{table}
    \centering
    \caption{Dataset statistics for 2-class classification.}
    \label{tab:2-class_split}
    \begin{tabular}{lcccc}
    \toprule
    \textbf{Pathology} & \textbf{Train} & \textbf{Valid} & \textbf{Test} & \textbf{Total} \\ \midrule
    Normal  & 3119 & 445 & 890 & 4454 \\
    COVID-19 & 3119 & 445 & 890 & 4454 \\ \midrule
    & 6238 & 890 & 1780 & 8908 \\
    \bottomrule
    \end{tabular}
\end{table}

\subsection{CoVScreen model}
The CoVScreen model has a CNN as backbone network tailored for the diagnosis of COVID-19 from chest X-Rays. The CoVScreen uses DenseNet~\cite{Huang:2017:DenseNet} as a base model with a dropout layer and two fully-connected layers added at the end of the network. The DenseNet has several compelling benefits including reducing the vanishing-gradient problem, reinforcing feature propagation, encouraging feature reuse, and reducing the number of parameters substantially. The CoVScreen model is based on \cite{singh:2024:computeraided}, applying DenseNet121 pre-trained on the ImageNet database~\cite{Deng:2009:ImageNet}. We use additional techniques including data pre-processing to improve data quality, data augmentation to improve model robustness to data variations, a class weighting scheme to handle class-imbalance problem, transfer learning to take advantage of learning low-level features from large-scale ImageNet dataset~\cite{Deng:2009:ImageNet}, and model fine-tuning to learn features specific to medical images. Application of these techniques helps model's generalisation ability as well as improving the model's robustness to noise, artifacts, and data variability. The techniques applied to tailor backbone CNN model to form CoVScreen model are described below:

\begin{itemize}
    \item Re-scaling images: We curated dataset from multiple sources where chest X-rays are acquired from different devices, different image acquisition parameters, and different image sizes. In order to have consistency and to feed images to the DenseNet121 backbone network, we resized all images to $224 \times 224$ pixels.
    
    \item Normalising images: Since the chest X-rays in our curated data are collected from multiple sites across the globe and do not follow the same acquisition protocol, the pixel intensity of each chest X-ray can vary considerably. We normalise the intensity value within the range $[0, 1]$. The normalisation step makes model less sensitive to small changes in weights and also helps in model convergence. 
    
    \item Data augmentation: In order to overcome over-fitting problem caused by small number of training images, we applied different image augmentation techniques including \emph{rotation}, \emph{zoom}, \emph{scaling}, \emph{horizontal flipping}, and \emph{shearing}. 
    
    \item Class weighting: Most of the medical imaging datasets have skewed distribution of samples for different classes, called \emph{data imbalance} problem in machine learning. Since machine learning models don't show robust performance on imbalanced datasets, it is important to give different weights to different classes depending upon their size in the dataset. We alter the weights of each class while computing loss during training phase in order to penalise model for any misclassification for the minority class. This is achieved by setting classes with minority samples to hold more weight and reducing weight for the majority classes. This \emph{class weighting} scheme affects the loss function by assigning class weights inversely proportional to their respective frequencies. The weights for each class are computed as follows:
    \begin{equation}
        w(c) = C_c \cdot \frac{\sum_{c=0}^N n_c}{N\cdot n_c}
    \end{equation}

    where $C_c$ is the class constant for a class $c$, $N$ is the number of classes, and $n_c$ is the number of samples in a class $c$. The computed class weights are used in the objective function (a.k.a loss function) to heavily penalise the false predictions concerned with the minority samples. 

    \item Transfer learning: Training deep learning model from scratch often require large-scale annotated data as well as GPU-scale computation and memory resources. The expert annotation of medical imaging data is expensive, time-consuming, and requires significant expertise. It is important to build a model which is already trained on large-scale annotated datasets. \emph{Transfer learning}, an alternative to model training from scratch, fine-tunes a CNN already pre-trained on a large-scale annotated dataset such as the ImageNet. This helps in speeding up model convergence while lowering computational complexity during training. In this study, we fine-tune the CoVScreen to learn features associated specifically to medical images. Thus, the model weights are tuned from generic feature maps to features associated with the chest X-ray dataset.
\end{itemize}

\section{Experimental Setup}\label{experimental_setup_sectioni}
All the experiments are implemented on a system having Ubuntu 18.04.1 LTS 64-bit operating system and Intel® Core™ i7-7700HQ CPU @ 2.80GHz × 8 processor. All programs are written in Python, where the software stack consists of PyTorch~\cite{Paszke:2019:PyTorch}, scikit-learn~\cite{scikit-learn}, and Keras~\cite{chollet2015keras} with the Tensorflow~\cite{Abadi:2016:Tensorflow} backend. The experiments were executed on 2 NVIDIA GTX 1080Ti graphics processing units (GPUs) having 11GB of on-board RAM with CUDA and CuDNN enabled to make the overall processing faster.

The images are scaled to $224 \times 224$ before being fed to the CoVScreen model. Also, images are converted to \emph{RGB} colour and normalised based on the mean and standard deviation. We remove the top layers of the model and freeze the previous convolutional layers. Then we add dense layers on the top of the remaining layers. We make use of dropout as a regularisation technique to prevent model from overfitting. For binary classification, \emph{categorical cross entropy loss} and \emph{softmax} activation function is used. The \emph{Adam} optimiser is used for training model in different experimental settings. We set various hyperparameters during training based on the grid search approach having \emph{learning rate}, $\eta$ of $1e-6$, \emph{number of epochs} of 30, and \emph{batch size} of 8. Early stopping is used to get best model performance in case model performance does not improve over the last three epochs. In this study, all experiments are performed by $5$-fold cross-validation, which is based on dividing all the data into $5$ folds, using single fold for testing and remaining folds for training and validation. $10\%$ of images for each class from training set are assigned as the validation set. The training step is repeated $5$ times until all folds are used for the test set.  To tackle class imbalance problem, we apply \emph{class weighting} scheme to penalise a model when it miss-classifies sample from the minority class. Furthermore, data augmentation is applied to add diversity in the training data by leveraging different transforms including \emph{horizontal flip}, \emph{rotation}, \emph{Gaussian noise}, and \emph{scaling}. During training, images are normalised as per the \emph{ImageNet} dataset specifications in order to use pre-trained CNN models with $mean = [0.485, 0.456, 0.406]$ and $std = [0.229, 0.224, 0.225]$. The network output layer provides scores for each of the $n$ classes, which are the probability scores in the range $[0-1]$ using the \emph{softmax} activation function. The final predicted class is selected based on the highest of the $n$ probability estimates obtained for the predicted classes. 

To evaluate effectiveness of the CoVScreen model, we report results in terms of accuracy score.  Although accuracy is an intuitive evaluation criterion for many classification problems, it is most suitable for balanced datasets. Given the imbalanced class scenario, with widely varying class distributions between different classes, we report precision, recall, F1-score, and and area under the ROC curve (AUC) score. 

\section{Results}\label{results_section}
This section provides experimental results under different experimental settings. We divided data into training, validation, and testing set having 70\%/10\%/20\% split. After this, we report results for each fold using 5-fold cross-validation. Finally, the mean and standard deviation for each metric is reported.

\subsection{Binary classification results separating Normal and COVID-19 cases}
To check the effectiveness of the CoVScreen model, we report 5-fold cross-validation results on the test set. Table~\ref{tab:results_binClf_4454_CXRs} shows performance for each fold in terms of precision, recall, f1-score, accuracy, and AUC scores. Although we have also calculated balanced accuracy score, given that number of Normal and COVID-19 chest X-rays are equal in this case, the balanced accuracy score is equal to the accuracy score. Based on these results, it is evident that the proposed model performance is quite low as compared to reported performance in the literature. 

\begin{table*}
  \centering
  \caption{Results for binary classification(Normal Vs. COVID-19) on the curated dataset.}
  \label{tab:results_binClf_4454_CXRs}
  \begin{tabular}{cccccc}
  \toprule
    \textbf{Fold} & \textbf{Precision} & \textbf{Recall} & \textbf{F1-score} & \textbf{Accuracy} & \textbf{AUC} \\ \hline
    Fold 1 &  0.7421 & 0.7432 & 0.7320 & 0.7432 & 0.8142 \\
    Fold 2 &  0.7351 & 0.6960 & 0.6828 & 0.6960 & 0.7929 \\
    Fold 3 &  0.7650 & 0.7275 & 0.7175 & 0.7275 & 0.8542 \\
    Fold 4 &  0.7497 & 0.7494 & 0.7493 & 0.7494 & 0.7974 \\
    Fold 5 &  0.7769 & 0.7769 & 0.7769 & 0.7769 & 0.8174 \\ 
    \hline
    Mean$\pm$Std & $0.7538\pm0.0176$ & $0.7386\pm0.0307$ & $0.7317\pm0.0362$ & $0.7386\pm0.0307$ & $0.8152\pm0.0249$ \\ 
    \midrule
  \bottomrule
  \end{tabular}
\end{table*}

To investigate the reason and to validate our hypotheses given in the contributions section, we carried out additional experiments in next sections.

\subsection{Impact on the diagnostic accuracy based on the COVID-19 severity}
In order to investigate the impact of severity of COVID-19 infection on the model performance, we conduct experiments by separating cases based on the COVID-19 severity. We analyse samples from \emph{COVIDGR1.0} dataset having samples categorised into \emph{normal-pcr+}, \emph{mild}, \emph{moderate}, and \emph{severe} categories. We run experiments in four settings depending on the four levels of COVID-19 severity. Table~\ref{tab:severity_analysis_dataset} provides number of chest X-rays for each of the four settings.

\begin{table*}
  \centering
  \caption{Data statistics for COVID-19 severity analysis.}
  \label{tab:severity_analysis_dataset}
  \begin{tabular}{lccc}
  \toprule
     \textbf{Setting 1: Normal vs. Normal-PCR+ cases} & & &  \\ \midrule
            & Train & Val & Test  \\ 
    Normal & 55 & 7 & 14 \\ 
    Normal-PCR+ & 55 & 7 & 14 \\ \hline
    \textbf{Setting 2: Normal vs. Mild cases} & & &  \\ \midrule
            & Train & Val & Test  \\ 
    Normal & 70 & 10 & 20 \\ 
    Mild & 70 & 10 & 20 \\ \hline
    \textbf{Setting 3: Normal vs. Moderate cases} & & & \\ \midrule
            & Train & Val & Test  \\ 
        Normal & 120 & 17 & 34 \\ 
        Moderate & 120 & 17 & 34 \\ \hline
    \textbf{Setting 4: Normal vs. Severe cases} & & & \\ \midrule
         & Train & Val & Test  \\ 
    Normal & 56 & 7 & 16 \\ 
    Severe & 56 & 7 & 16 \\ \bottomrule
  \end{tabular}
\end{table*}

Based on the experimental results in Table~\ref{tab:results_binClf_severity}, we find that the diagnostic accuracy of the model is highly dependent on the severity level of the COVID-19 infection. For the \emph{Normal-PCR+} class, where the chest X-rays do not show any visual features of COVID-19 but the actual RT-PCR test is positive, the performance of the model in separating \emph{Normal} cases from \emph{Normal-PCR+} is very poor. The low performance of the model is justified given no visual features separating the two classes are present in the chest X-rays. For \emph{mild} cases, where radiographic features are not fully developed, it is challenging for the model to separate mild COVID-19 cases from healthy cases. On the other hand, for severe cases, where radiographic features are clearly visible on the chest X-rays, it is easy for the model to distinguish healthy and severe COVID-19 chest X-rays. Several other studies~\cite{Kundu:2020:how_might_AI_and_chest_imaging,Tabik:2020:COVIDGR} have also reported variation in the model performance depending on the severity of cases, indicating model reliance on the dataset as well as the severity of the cases in the dataset. 

\begin{table*}
  \centering
  \caption{Results for diagnosing COVID-19 cases as per their severity level.}
  \label{tab:results_binClf_severity}
  \begin{tabular}{cccccc}
  \toprule
  \multicolumn{6}{@{}l}{\textbf{Setting 1: Normal PCR+ vs. Negative}} \\ \hline
    \textbf{Fold} & \textbf{Precision} & \textbf{Recall} & \textbf{F1-score} & \textbf{Accuracy} & \textbf{AUC} \\ \hline
    Fold 1 & 0.4333 & 0.4330 & 0.4327 & 0.4333 & 0.4955  \\
    Fold 2 & 0.3500 & 0.3666 & 0.3485 & 0.3666 & 0.4311 \\
    Fold 3 & 0.4305 & 0.4333 & 0.4276 & 0.4333 & 0.4266 \\
    Fold 4 & 0.4074 & 0.4666 & 0.3650 & 0.4666 & 0.4955 \\
    Fold 5 & 0.4305 & 0.4333 & 0.4276 & 0.4333 & 0.4466 \\ 
    \hline
    Mean$\pm$Std & $0.4103\pm0.0364$ &  $0.4266\pm0.0376$ & $0.4003\pm0.0414$ & $0.4266\pm0.0376$ & $0.4591\pm0.0351$\\ 
    \midrule
    \multicolumn{6}{@{}l}{\textbf{Setting 2: Mild vs. Negative}} \\ \hline
    Fold 1 & 0.6010 & 0.6000 & 0.5989 & 0.6000 & 0.5450 \\
    Fold 2 & 0.8030 & 0.6750 & 0.6366 & 0.6750 & 0.7200 \\
    Fold 3 & 0.6648 & 0.6500 & 0.6419 & 0.6500 & 0.6662 \\
    Fold 4 & 0.6278 & 0.6250 & 0.6228 & 0.6250 & 0.5837 \\
    Fold 5 & 0.5854 & 0.5750 & 0.5615 & 0.5750 & 0.5915 \\
     \hline
    Mean$\pm$Std & $0.6564\pm0.0900$ & $0.6250\pm0.0407$ & $0.6123\pm0.0339$ & $0.6250\pm0.0407$ & $0.6213\pm0.0726$ \\ 
    \midrule
     \multicolumn{6}{@{}l}{\textbf{Setting 3: Moderate vs. Negative}} \\ \hline
    Fold 1 &  0.7502 & 0.7500 & 0.7499 & 0.7500 & 0.7655 \\
    Fold 2 &  0.7688 & 0.7500 & 0.7455 & 0.7500 & 0.8209 \\
    Fold 3 &  0.7684 & 0.7647 & 0.7638 & 0.7647 & 0.7599 \\
    Fold 4 &  0.7125 & 0.7058 & 0.7035 & 0.7058 & 0.6898 \\
    Fold 5 &  0.7519 & 0.7500 & 0.7495 & 0.7500 & 0.7716 \\
     \hline
    Mean$\pm$Std & $0.7504\pm0.0236$ & $0.7441\pm0.0230$ & $0.7424\pm0.0235$ & $0.7441\pm0.0230$ & $0.7615\pm0.0483$ \\ 
    \midrule
    \multicolumn{6}{@{}l}{\textbf{Setting 4: Severe vs. Negative}} \\ \hline
    Fold 1 &  0.8333 & 0.8125 & 0.8095 & 0.8125 & 0.8691 \\
    Fold 2 &  0.8562 & 0.8437 & 0.8423 & 0.8437 & 0.8417 \\
    Fold 3 &  0.8478 & 0.7812 & 0.7702 & 0.7812 & 0.8867 \\
    Fold 4 &  0.7914 & 0.7812 & 0.7793 & 0.7812 & 0.9003 \\
    Fold 5 &  0.8750 & 0.8333 & 0.8285 & 0.8333 & 0.9066 \\
     \hline
    Mean$\pm$Std &  $0.8407\pm0.0324$ & $0.8104\pm0.0298$ & $0.8060\pm0.0319$ & $0.8104\pm0.0298$ & $0.8809\pm0.0270$\\ 
  \bottomrule
  \end{tabular}
\end{table*}

\subsection{Impact on the diagnostic accuracy based on the paediatric patients' chest X-rays}
\begin{figure*}
  \setcounter{subfigure}{0}
  \centering
  \subfigure[Normal (paediatric patient)]{\includegraphics[width=3.5cm, height=3cm]{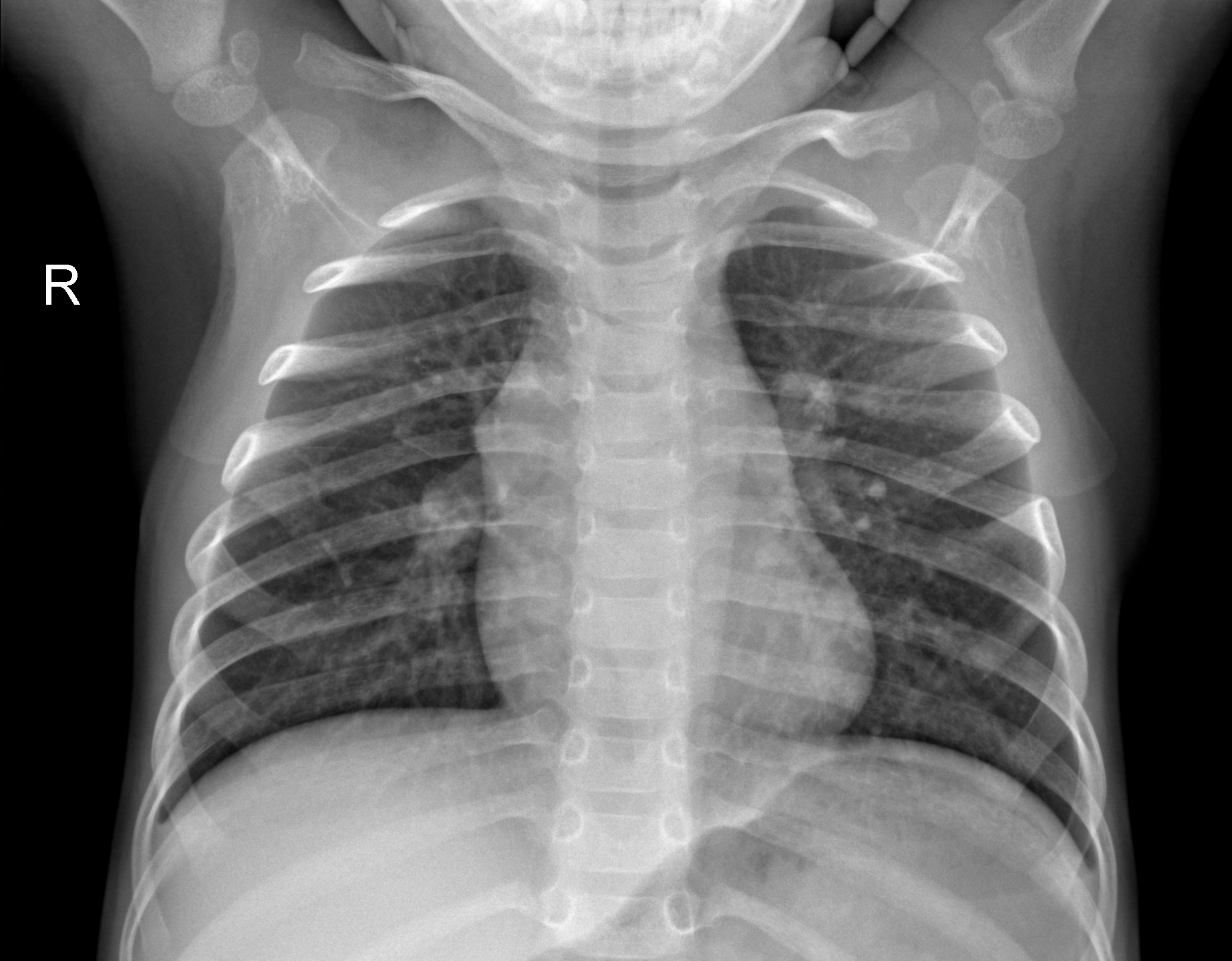}}\quad
  \subfigure[Bacterial pneumonia (paediatric patient)]{\includegraphics[width=3.5cm, height=3cm]{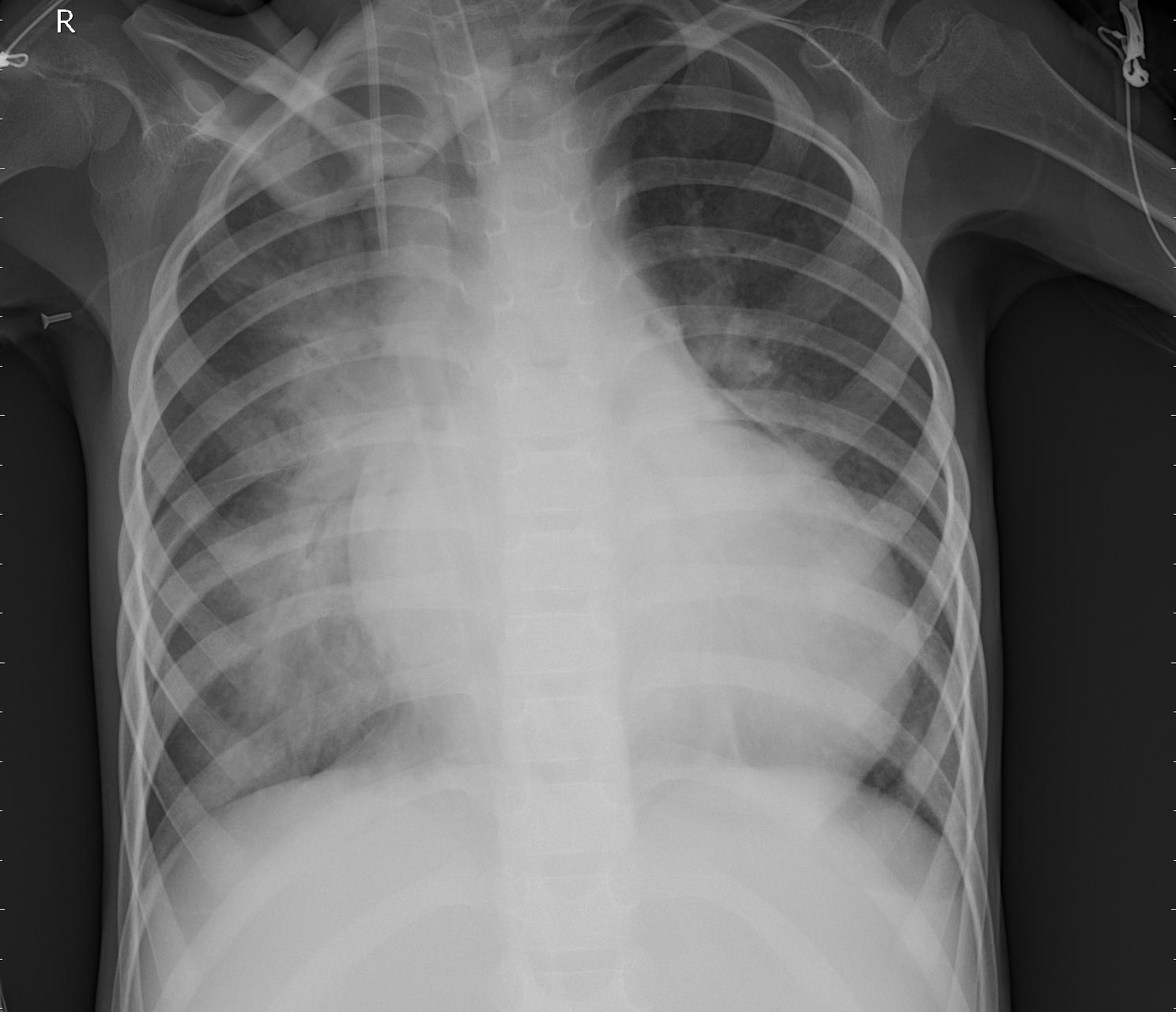}}\quad
  \subfigure[Viral pneumonia (paediatric patient)]{\includegraphics[width=3.5cm, height=3cm]{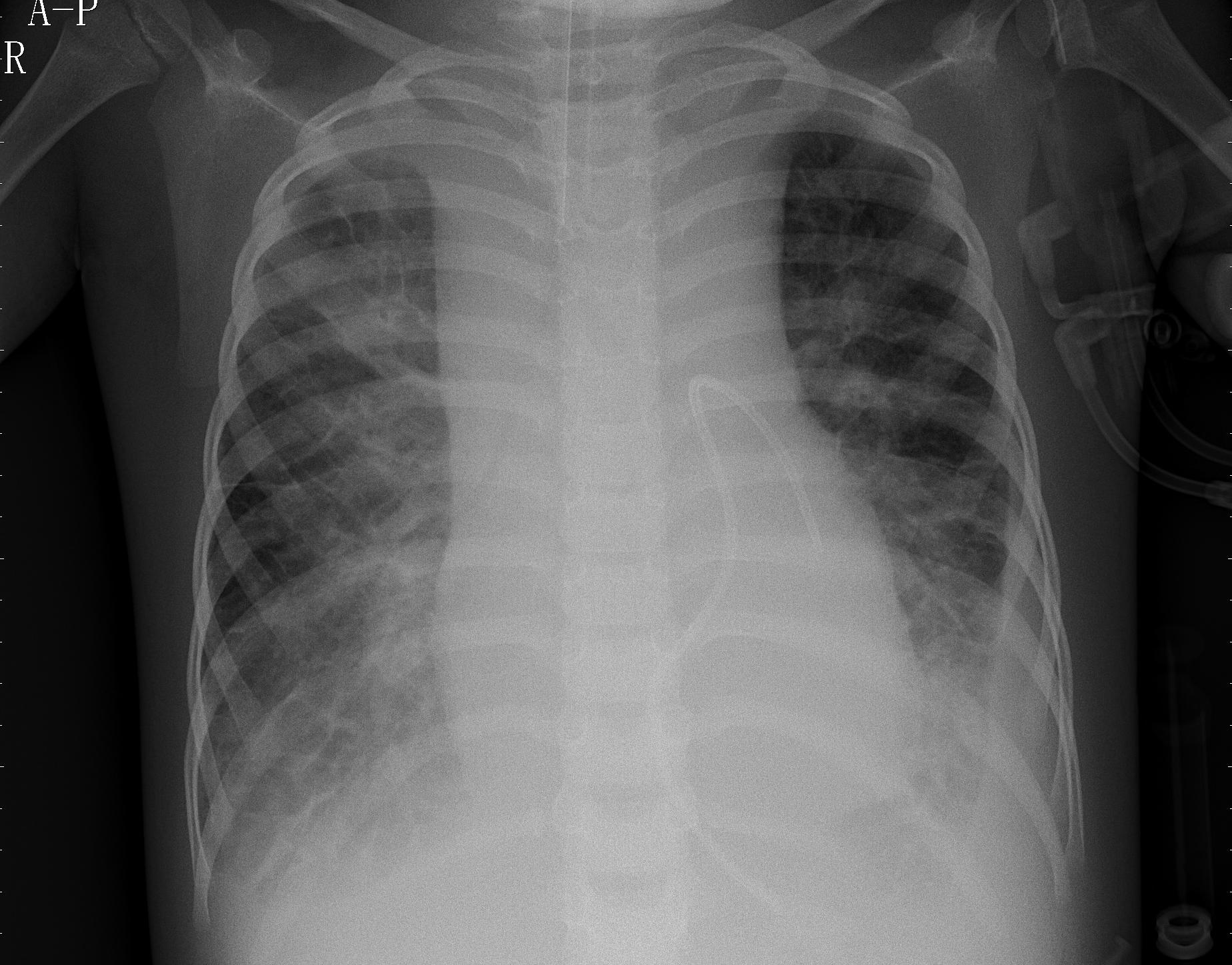}}\quad
  \subfigure[COVID-19 (Adult)]{\includegraphics[width=3.5cm, height=3cm]{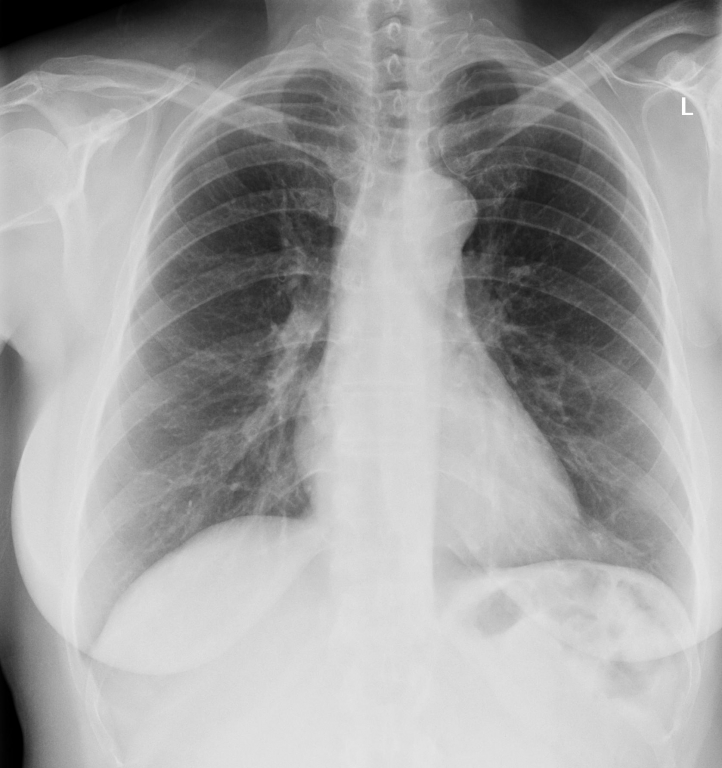}}\quad \\
  \caption{Differences in chest X-rays from paediatric patients vs. COVID-19 chest X-ray from adult population.} 
  \label{fig:pediatricVsCOVID19}
\end{figure*}

As highlighted in related work section~\ref{related_work_section}, we found that many studies get \emph{healthy} and \emph{pneumonia} chest X-rays from the ``Paul Mooney Pneumonia Detection Challenge" hosted at \emph{Kaggle}~\cite{kaggle:paul_mooney_chest_X-ray,Kermany:Goldbaum:2018_identifying_medical_diagnosis}. This dataset have chest X-rays which were selected from retrospective cohorts of pediatric patients of one to five years old from Women and Children's Medical Center, Guangzhou, China. This dataset has three classes, namely, \emph{Normal}, \emph{Bacterial Pneumonia}, and \emph{Viral Pneumonia}. The \emph{normal} chest X-ray depicts clear lungs without any areas of abnormal opacification in the image. Bacterial pneumonia typically exhibits a focal lobar consolidation, whereas viral pneumonia manifests with a more diffuse ‘‘interstitial’’ pattern in both lungs. Figure~\ref{fig:pediatricVsCOVID19} shows random sample of each of the three classes and a random sample from the curated COVID-19 dataset. 

\begin{figure*}
  \setcounter{subfigure}{0}
  \centering
  \subfigure[Adult chest X-ray showing arched ribs]{\includegraphics[width=3.5cm, height=3cm]{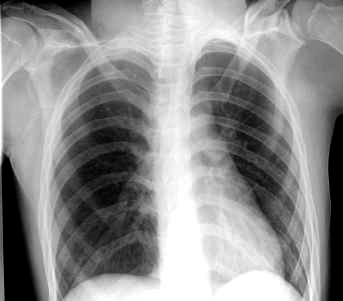}}\qquad
  \subfigure[Neonate chest X-ray showing flattened ribs]{\includegraphics[width=3.5cm, height=3cm]{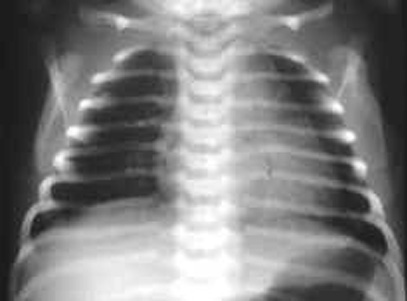}} \\
  \caption{Differences in ribs positioning visible in chest X-rays from paediatric patients vs. chest X-ray from adult population. Source: \url{https://www.rch.org.au/trauma-service/manual/how-are-children-different/}} 
  \label{fig:ribs_positioning}
\end{figure*}

From a radiologist's perspective, there are key differences between chest X-rays captured from paediatric patients and adult population. First, the size of children's chest is small compared to adult's chest, which can also be visible in Figure~\ref{fig:pediatricVsCOVID19}. Second, the rib of children are positioned more horizontally than those of adults as can be seen in Figure~\ref{fig:ribs_positioning}. Given that children and infants differ, both anatomically and physiology, from adults, these differences have an impact on the clinical assessment using chest X-rays. A learning algorithm can learn these key differences between paediatric population and adult population based on the chest X-rays, leading to model bias as model can report results solely based on these differences rather than radiographic features of a particular disease or abnormality. Hence, we hypothesise not to train model having chest X-rays from mix of both children and adult population. To prove our hypothesis, we run additional experiments to classify COVID-19 vs. Normal (healthy) cases by taking \emph{normal} cases from different sources. Table~\ref{tab:two_settings_distribution} shows distribution of \emph{normal} and \emph{COVID-19} cases for binary classification under two settings. For each of the two settings, we split data for each class into training (1106 images), validation (158 images), and testing (316) images, making in total 1580 chest X-rays.   

\begin{table*}
  \centering
  \caption{Data statistics for different settings of binary classification (Normal vs. COVID-19).}
  \label{tab:two_settings_distribution}
  \begin{tabular}{p{2cm}p{4cm}p{9cm}}
  \toprule
  Setting & COVID-19 & Normal \\ \midrule
  Setting 1 & 1580 (Curated dataset) & 1580 (all normal images taken from ChestX-ray14 (adult population))\\
  Setting 2 & 1580 (Curated dataset) & 1580 (all normal images taken from Kaggle pneumonia challenge by Paul Mooney (paediatric population))\\
\bottomrule
\end{tabular}
\end{table*}

\begin{table*}
  \centering
  \caption{Results of binary classification (Normal Vs. COVID-19) under different settings.}
  \label{tab:results_two_settings_binaryClassification}
  \begin{tabular}{cccccc}
  \toprule
  \multicolumn{6}{@{}l}{\textbf{Setting 1: Adult Normal CXRs vs. Adult COVID-19 CXRs}} \\ \hline
     \textbf{Fold} & \textbf{Precision} & \textbf{Recall} & \textbf{F1-score} & \textbf{Accuracy} & \textbf{AUC} \\ \hline
      Fold 1 & 0.8056 & 0.7547 & 0.7440 & 0.7547 & 0.8521 \\
      Fold 2 & 0.8204 & 0.8196 & 0.8195 & 0.8196 & 0.8581 \\
      Fold 3 & 0.7883 & 0.7325 & 0.7109 & 0.7325 & 0.8117 \\ 
      Fold 4 & 0.7617 & 0.7294 & 0.7208 & 0.7294 & 0.7816 \\
      Fold 5 & 0.7824 & 0.6977 & 0.6733 & 0.6977 & 0.8170 \\ \hline
      Mean$\pm$Std & $0.7917\pm0.0231$ & $0.7468\pm0.0469$ & $0.7337\pm0.056$ & $0.7468\pm0.0469$ & $0.8241\pm0.0324$ \\
  \midrule
   \multicolumn{6}{@{}l}{\textbf{Setting 2: Paediatric Normal CXRs vs. Adult COVID-19 CXRs}} \\ \hline
      \textbf{Fold} & \textbf{Precision} & \textbf{Recall} & \textbf{F1-score} & \textbf{Accuracy} & \textbf{AUC} \\ \hline
      Fold 1 & 0.9747 & 0.9746 & 0.9746 & 0.9746 & 0.9956 \\
      Fold 2 & 0.9678 & 0.9667 & 0.9667 & 0.9667 & 0.9940 \\
      Fold 3 & 0.9768 & 0.9762 & 0.9762 & 0.9762 & 0.9953 \\
      Fold 4 & 0.9713 & 0.9699 & 0.9699 & 0.9699 & 0.9981 \\
      Fold 5 & 0.9858 & 0.9857 & 0.9857 & 0.9857 & 0.9984 \\ \hline
      Mean$\pm$Std & $0.9753\pm0.0070$ & $0.9746\pm0.0074$ & $0.9746\pm0.0074$ & $0.9746\pm0.0074$ & $0.9963\pm0.0019$ \\
  \bottomrule
  \end{tabular}
\end{table*}

Table~\ref{tab:results_two_settings_binaryClassification} shows results for the binary classification (Normal CXRs vs. COVID-19 CXRs) under two different settings. For both of the experimental settings, we keep same COVID-19 chest X-rays but changes the source of normal chest X-rays. This is to ensure variation in results because of source of chest X-rays (paediatrics vs. adults). Based on results, we find that when chest X-rays for both the classes (Normal vs. COVID-19) are from adult population, the results in terms of different classification metrics are lower when compared to results got when COVID-19 chest X-rays are from adult population whereas normal chest X-rays are from paediatric patients. This indicates model bias towards anatomical differences between paediatric population and adult population rather than learning distinguishable radiographic features of a disease (or abnormality).

\subsection{Diagnostic accuracy comparing paediatric normal vs. adult normal chest radiographs}
In order to validate the hypothesis that machine learning models report biased results with a mix of population, we run experiment for \emph{binary classification} to distinguish adult and paediatric chest X-rays. For the paediatric population, we take \emph{normal} chest X-rays from \emph{Kaggle Paul Mooney pneumonia challenge}, giving us a total of $1583$ \emph{normal} (or healthy) chest X-rays. For the adult population, we take same number of \emph{normal} (No Findings) from Chest X-ray14 dataset. We split the entire data into training (70\%), validation (10\%), and testing (20\%). Table~\ref{tab:results_paediatricVsadult} shows results 5-fold cross-validation of binary classification separating \emph{normal paediatric chest X-rays } and \emph{normal adult chest X-rays}. We run experiment for $5$ runs and calculated \emph{mean} and \emph{standard deviation} across different classification metrics. Experimental results shows that model can distinguish normal paediatric chest X-rays from normal adult chest X-rays with \emph{precision} ($0.9781\pm0.0079$), \emph{recall}($0.9744\pm0.0091$), \emph{f1-score}($0.9744\pm0.0091$), \emph{accuracy}($0.9744\pm0.0091$), and \emph{AUC-ROC}($0.9975\pm0.0009$). Given that for both the classes, we have normal (No Findings) chest X-rays, model performance is quite good in classifying two classes, indicating that model is learning data characteristics which are specific to differences in chest X-rays for paediatric population and adult population. 

\begin{table*}
  \centering
  \caption{Results of binary classification (Normal paediatric CXRs vs. Normal adult CXRs)}
  \label{tab:results_paediatricVsadult}
  \begin{tabular}{cccccc}
  \toprule
 \textbf{Fold} & \textbf{Precision} & \textbf{Recall} & \textbf{F1-score} & \textbf{Accuracy} & \textbf{AUC} \\ \hline
  Fold 1 & 0.9802 & 0.9794 & 0.9794 & 0.9794 &  0.9977 \\ 
  Fold 2 & 0.9674 & 0.9651 & 0.9651 & 0.9651 &  0.9964 \\ 
  Fold 3 & 0.9765 & 0.9762 & 0.9762 & 0.9762 &  0.9967 \\ 
  Fold 4 & 0.9773 & 0.9762 & 0.9762 & 0.9762 &  0.9980 \\ 
  Fold 5 & 0.9889 & 0.9889 & 0.9899 & 0.9899 &  0.9986 \\ \hline
  Mean$\pm$Std & $0.9781\pm0.0079$ & $0.9744\pm0.0091$ & $0.9744\pm0.0091$ & $0.9744\pm0.0091$ & $0.9975\pm0.0009$ \\
  \bottomrule
  \end{tabular}
\end{table*}

The proposed methodology can further be extended to diagnose COVID-19 in \emph{multi-label settings}, where a particular chest X-ray can have more than one disease. We try to formulate dataset by combining ChestX-ray14 dataset with COVID-19 samples. However, we find that this is not valid given we don't have any samples where COVID-19 co-exist with any of the fourteen thoracic diseases present in the ChestX-ray14 dataset. This would bias the results given all samples having COVID-19 still be treated having single-label and can inflate results compared to other fourteen diseases which are in multi-label settings. Given limitation of not having appropriate dataset in a multi-label setting having COVID-19 as one of the diseases, we restrict this study to binary classification. 

\subsection{Looking at few qualitative examples}
\begin{figure*}
  \setcounter{subfigure}{0}
  \begin{tcolorbox}[colback=green!5!white,colframe=green!75!black,title=Correct predictions]
  \centering
  \subfigure[COVID-19 case correctly identified] {\includegraphics[width=14cm, height=2.8cm]{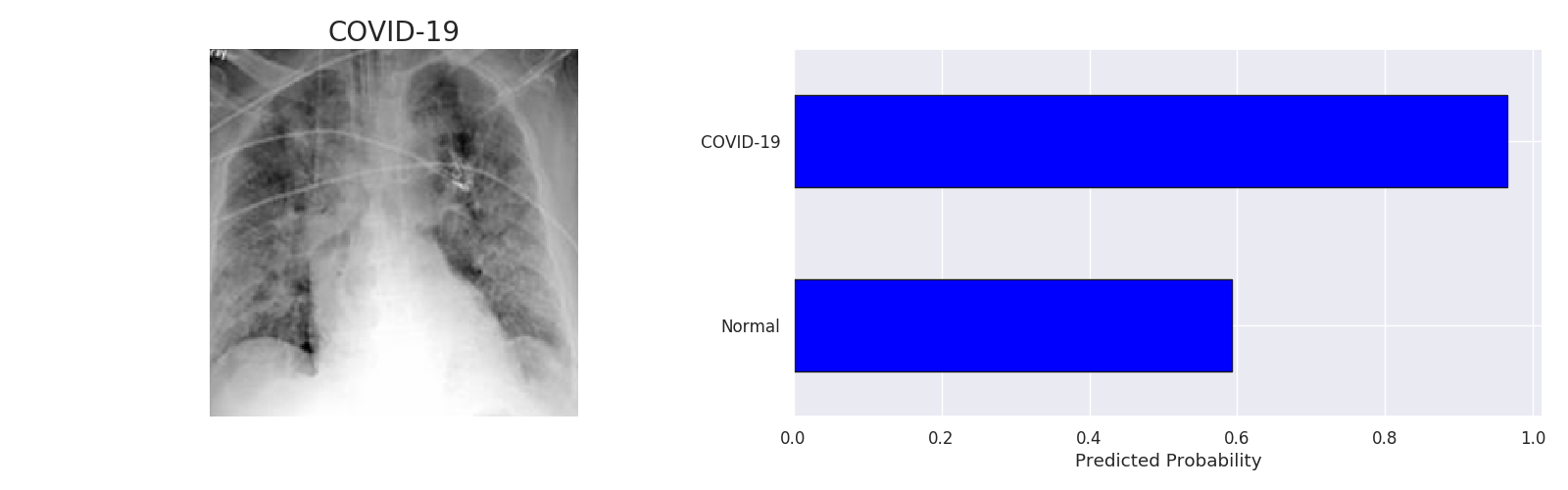}}  \\
  \subfigure[Paediatric patient correctly identified] {\includegraphics[width=14cm, height=2.8cm]{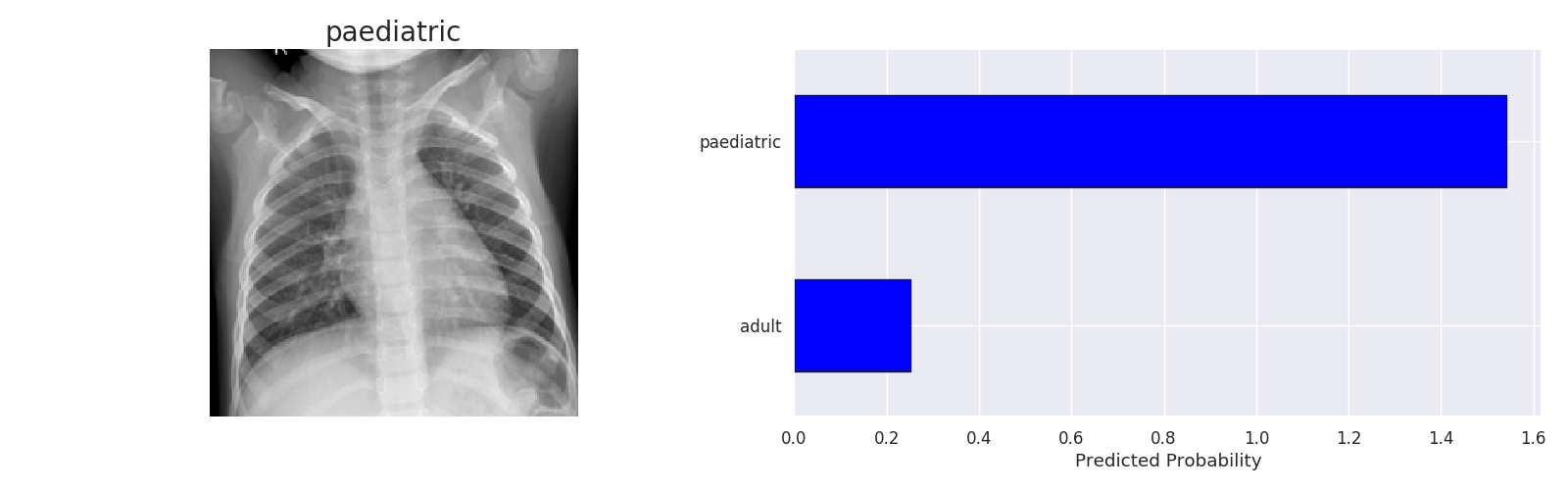}}  \\
  \subfigure[Adult patient correctly identified] {\includegraphics[width=14cm, height=2.8cm]{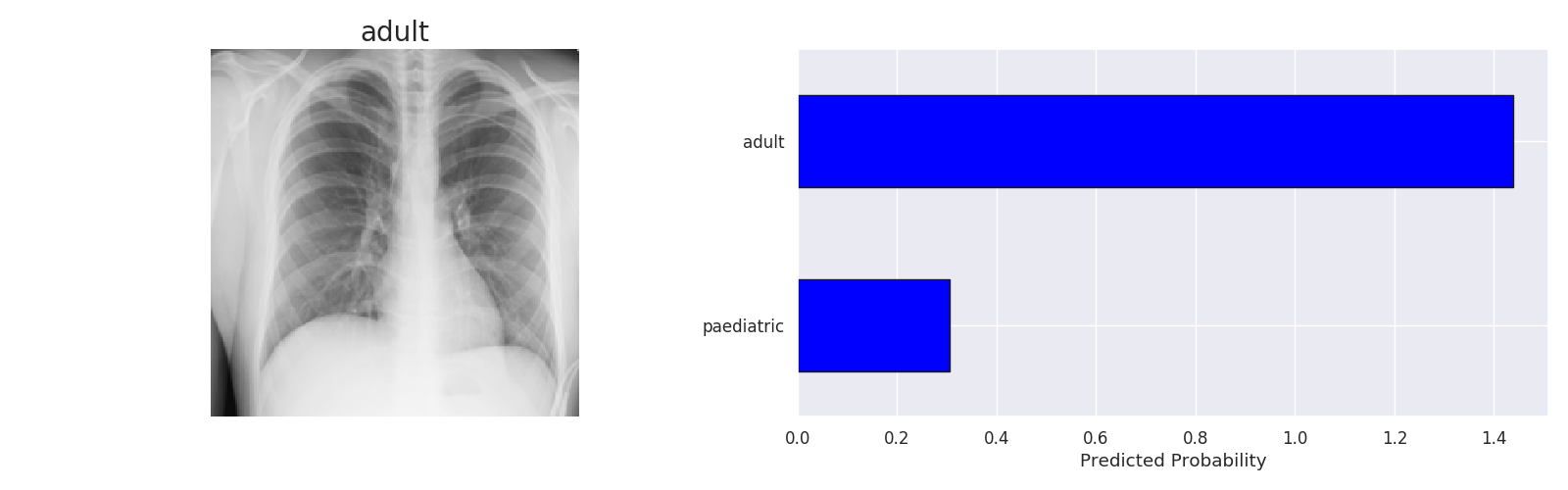}}  \\
  \subfigure[Mild case of COVID-19 correctly identified] {\includegraphics[width=14cm, height=2.8cm]{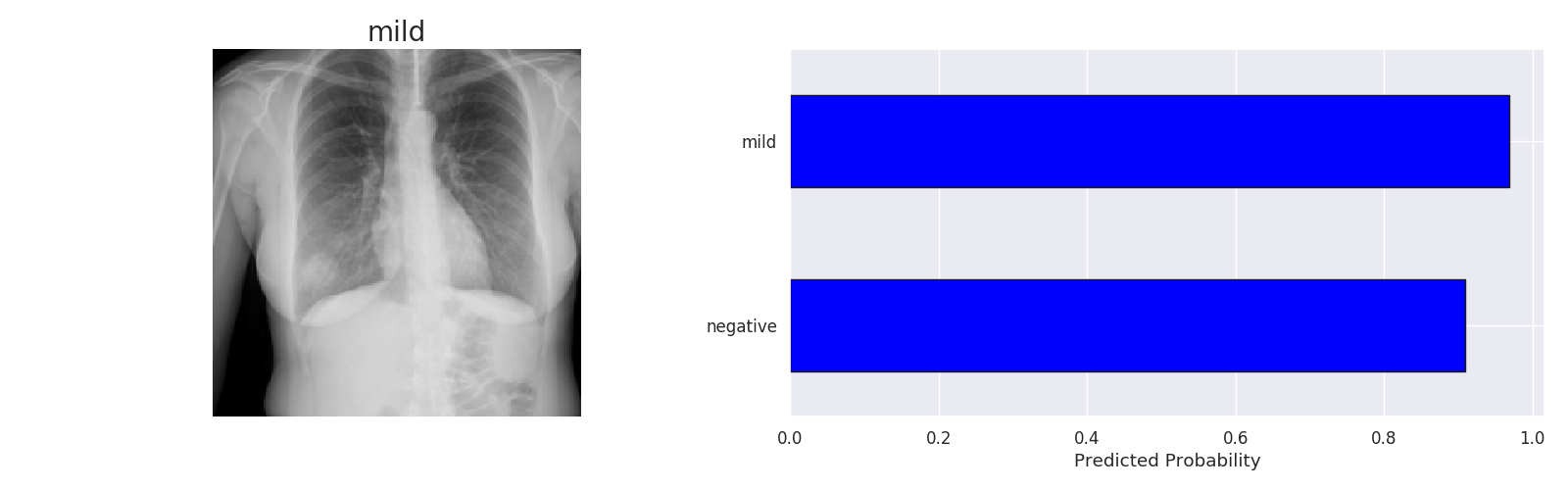}}  \\
  \subfigure[Moderate case of COVID-19 correctly identified]{\includegraphics[width=14cm, height=2.8cm]{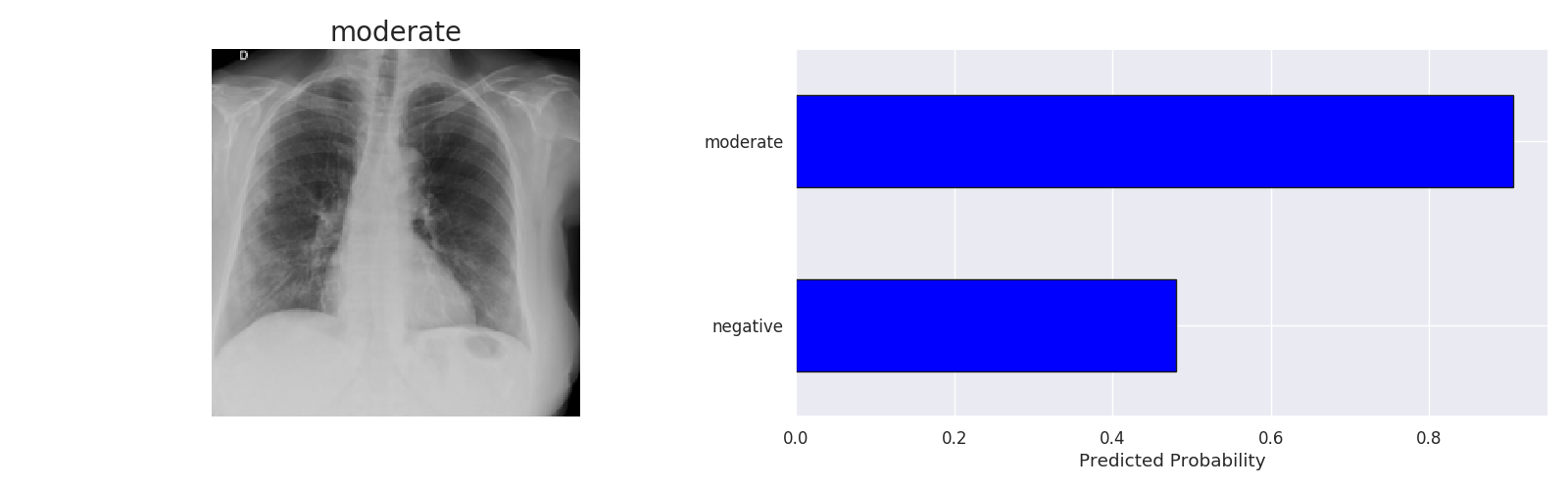}}  \\
  \subfigure[Severe case of COVID-19 correctly identified] {\includegraphics[width=14cm, height=2.8cm]{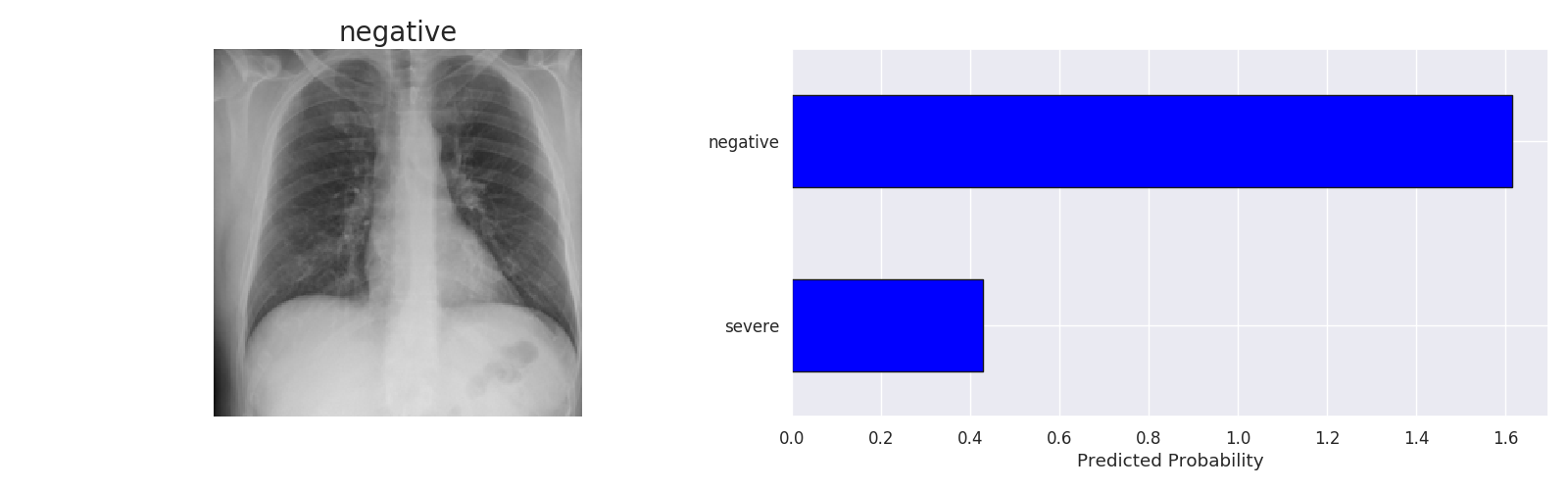}}  \\
  \end{tcolorbox}
  \caption{Correct predictions by model selected randomly under different classification settings.}
  \label{fig:correct-predictions}
\end{figure*}

\begin{figure*}
\setcounter{subfigure}{0}
  \begin{tcolorbox}[colback=red!5!white,colframe=red!75!black,title=Wrong predictions]
  \centering
\subfigure[COVID-19 case misclassified as Normal] {\includegraphics[width=14cm, height=2.8cm]{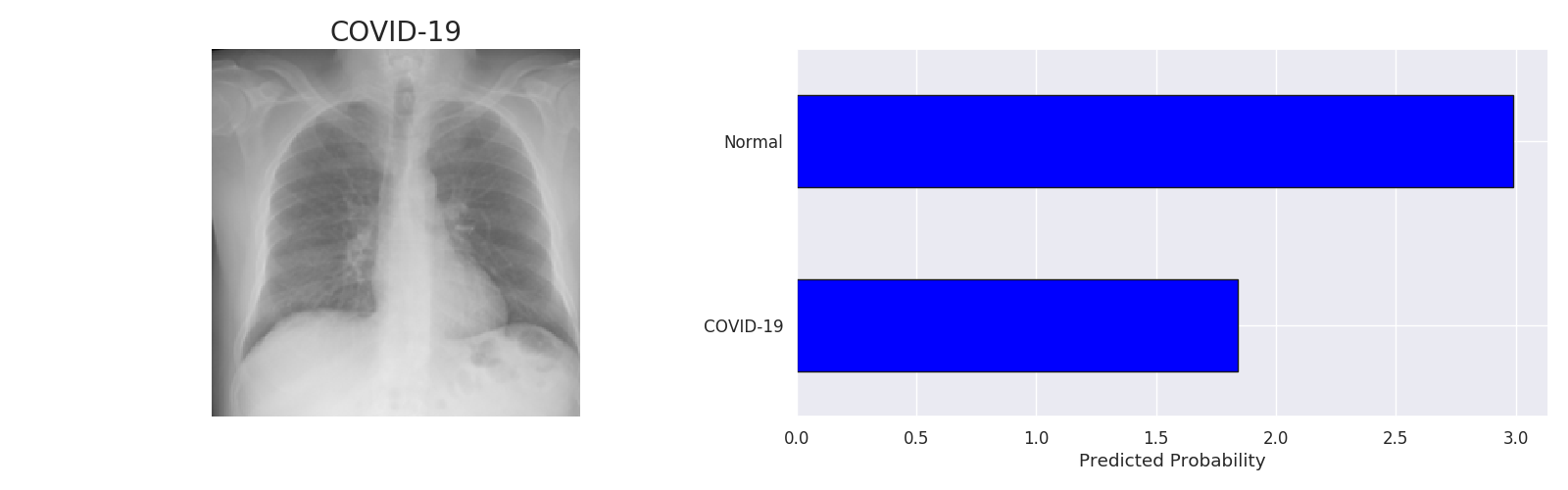}}  \\
  \subfigure[Paediatric case misclassified as Adult] {\includegraphics[width=14cm, height=2.8cm]{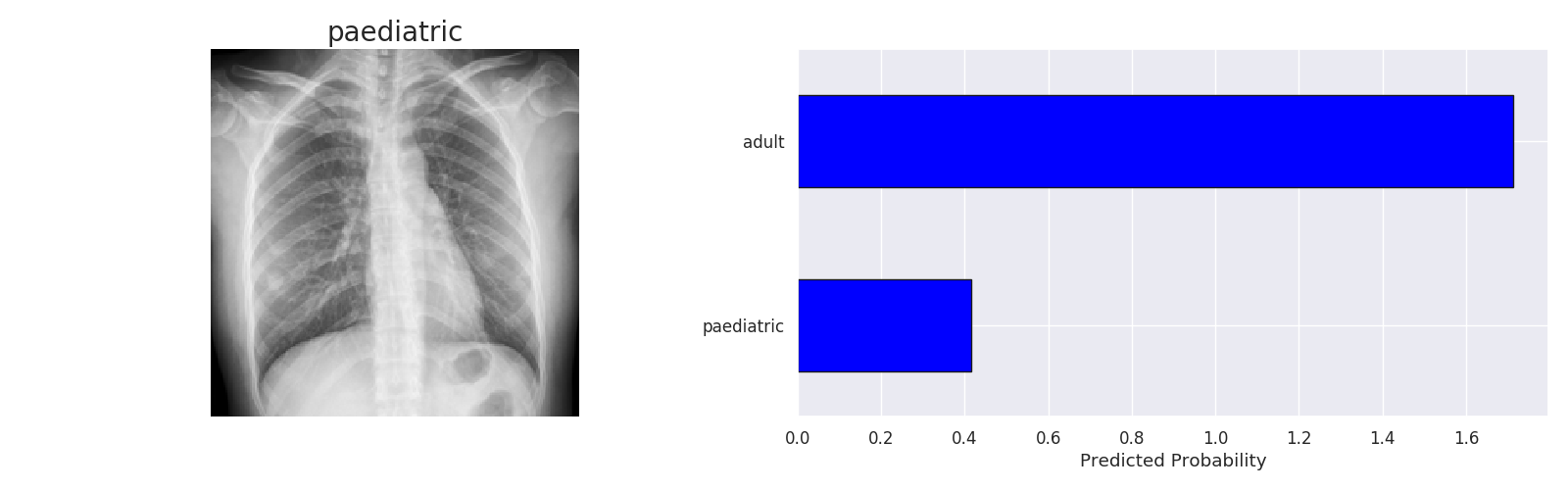}}  \\
  \subfigure[Negative case misclassified as Normal-PCR+] {\includegraphics[width=14cm, height=2.8cm]{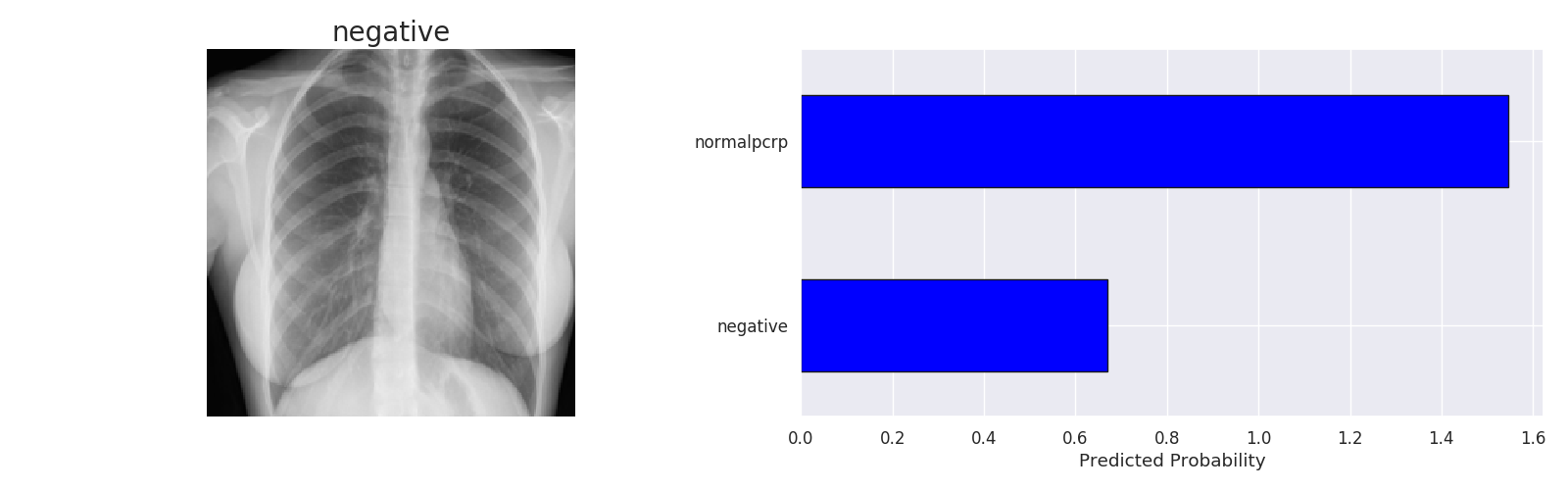}}  \\
  \subfigure[Negative case misclassified as mild COVID-19]{\includegraphics[width=14cm, height=2.8cm]{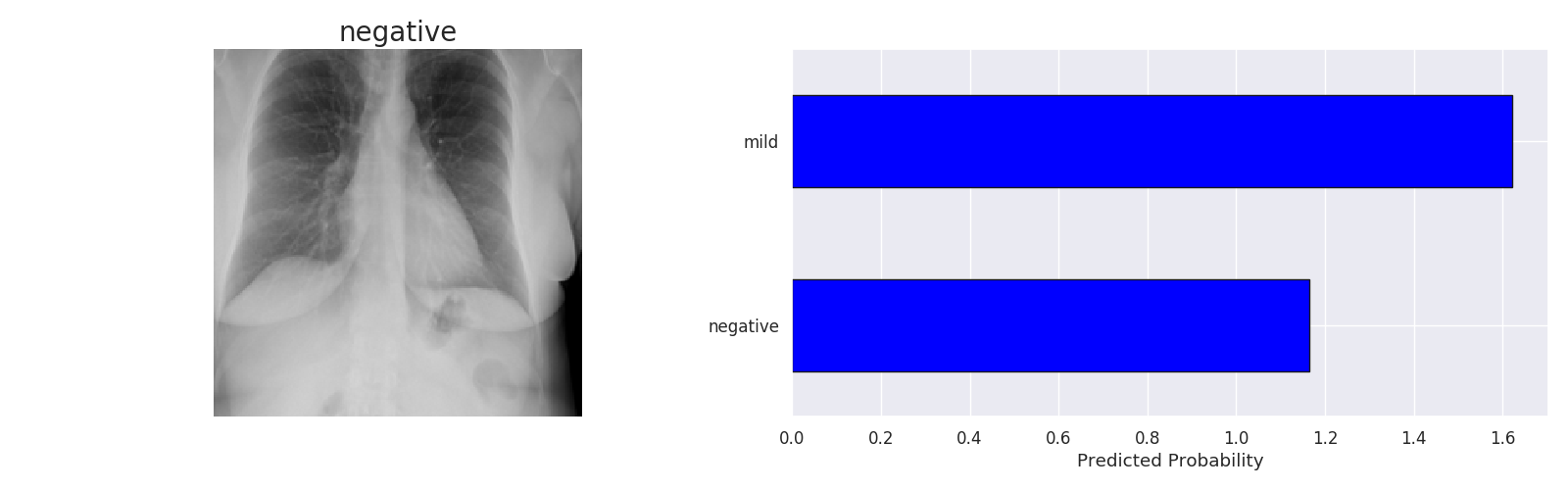}}  \\
  \subfigure[Mild COVID-19 case misclassified as Negative] {\includegraphics[width=14cm, height=2.8cm]{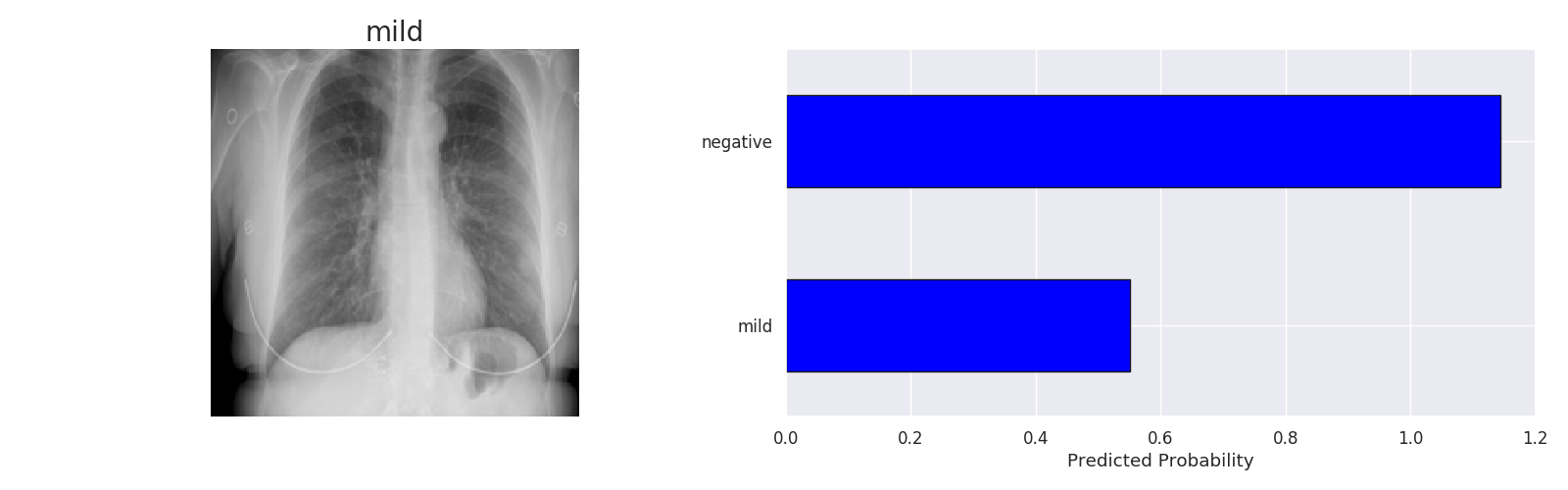}}  \\
  \subfigure[Severe COVID-19 case misclassified as Negative] {\includegraphics[width=14cm, height=2.8cm]{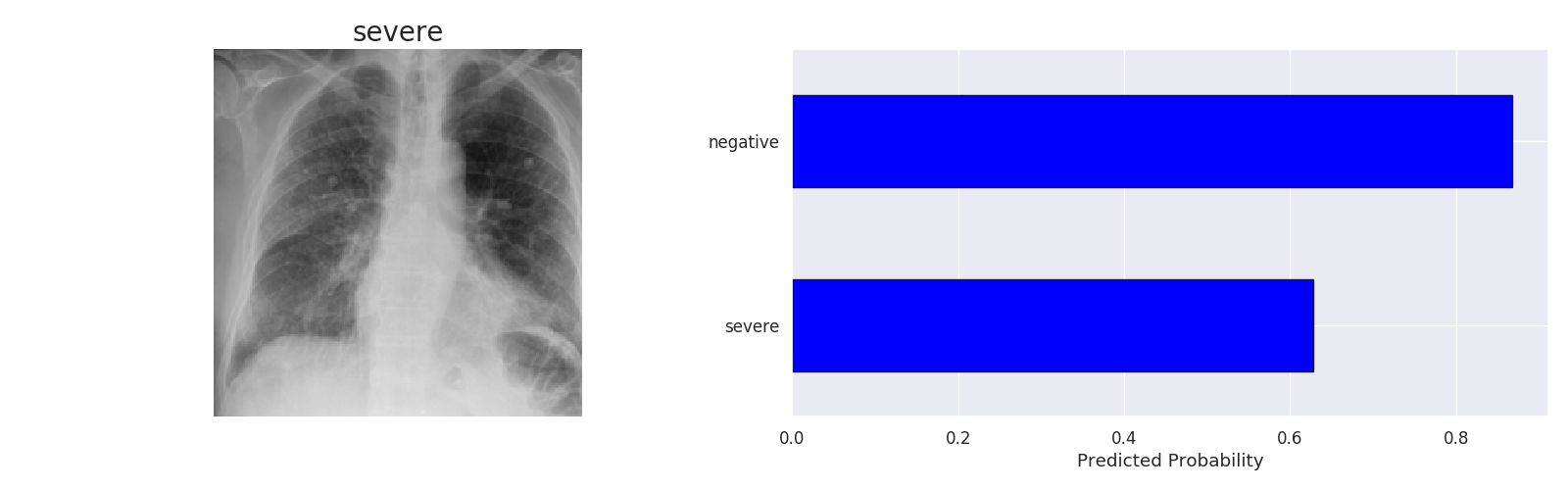}}  \
  \end{tcolorbox}
  \caption{Wrong predictions by model selected randomly under different classification settings.}
  \label{fig:wrong-predictions}
\end{figure*}

To see for what images the model is making correct predictions and for what images model is mis-classifying, we show here randomly selected few examples. Figure~\ref{fig:correct-predictions} shows sample examples of correct predictions made by the model under different experimental settings. Figure~\ref{fig:wrong-predictions} shows sample examples where model fails to predict the correct class of the input image. A closer look to these examples reflect that model is doing well in distinguishing paediatric chest X-rays and adult chest X-rays because of distinguishable anatomical features. Also, model is doing correct predictions for COVID-19 cases where the severity of COVID-19 infection is high. On the other hand, model is failing for mild and moderate cases where the radiographic features of COVID-19 infection are not fully developed and not visible in chest X-rays.

\section{Comparison with existing work}
In this section we compare our curated dataset with datasets in related studies. The direct comparison of results is not possible because of lack of a benchmark dataset. Also, given most of the studies have taken chest X-rays from databases that have been continuously updated, it is challenging to get same set of images used in existing studies. Table~\ref{tab:comparisonWExistingWork} shows comparison with few of the representative studies indicating that most studies have limited number of COVID-19 samples, ranging from 25 samples to few hundreds. In this study, we curated large-scale dataset from multiple databases, giving 4,454 COVID-19 infected chest X-rays. 

Due to the database updates over time and the public availability of other data collections, it is impossible to carry out direct comparison of model results. Table~\ref{tab:comparison_results} shows comparison of current study with some representative studies having similar dataset settings. For comparison, we use \emph{Normal} and \emph{COVID-19} chest X-rays from a mix of population having both paediatric and adult population, similar to datasets curated in representative studies. Table~\ref{tab:comparison_results} shows that the results from the present study are consistent with results from previous COVID-19 studies. These results shows model's effectiveness in distinguishing COVID-19 infection cases from most of the common thoracic diseases. However, as discussed and results shown in Section~\ref{results_section}, we highlighted that various factors such as limited dataset size, standardisation of data from different sources, class imbalance, age of subjects (mix of paediatric and adult population) in a study, as well as the severity of the disease, impact the classification performance of the model. The main goal of this study is to outline thoughtful dataset curation and evaluation of classification performance. 

\begin{table*}
    \centering
     \caption{Comparison of our proposed work with the existing work in terms of number of COVID-19 chest X-rays.}
    \label{tab:comparisonWExistingWork}
    \begin{tabular}{p{8cm}p{4cm}}
    \toprule
    \textbf{Study}  & \textbf{\# COVID-19} \\
    \midrule
    Wang and Wong~\cite{wang:2020:covidnet} & 358 \\
    Sethy and Behera~\cite{sethy:2020:covid-19} & 127 \\
    Hemdan \textit{et al.}~\cite{hemdan:2020:covidxnet} & 25\\
    Narin \textit{et al.}~\cite{narin:2020:automatic} & 341 \\
    Ozturk \textit{et al.}~\cite{ozturk:2020:DarkNet} & 127 \\
    Khan \textit{et al.}~\cite{khan:2020:CoroNet} & 284 \\
    Hussain \textit{et al.}~\cite{Hussain:Parvez:2020:CoroDet} & 500 \\
    \textbf{This study} & \textbf{4454} \\
    \bottomrule
    \end{tabular}
\end{table*}

\begin{table*}[]
    \centering
    \caption{Comparison of our proposed work with the existing work in terms of diagnostic performance for COVID-19.}
    \label{tab:comparison_results}
    \begin{tabular}{p{3cm}p{2.5cm}p{6cm}p{3.5cm}}
         \toprule
        \textbf{Study} & \textbf{Model} & \textbf{Dataset} & \textbf{Results} \\
         \midrule
         Wang and Wong~\cite{wang:2020:covidnet} & COVID-Net pretrained on the ImageNet & 5,941 chest X-rays across 2839 patients (1203 normal + 45 COVID-19 + 660 non-COVID viral pneumonia + 931 bacterial pneumonia) & Accuracy (binary classification): 92.4\% and Accuracy (4-class classification): 83.5\% \\
         Sethy and Behera~\cite{sethy:2020:covid-19} & Deep features from ResNet-50 + SVM classifier & Dataset collected from Github and Kaggle~\cite{kaggle:paul_mooney_chest_X-ray} comprising 50 chest X-rays (25 COVID-19 cases and 25 normal cases) & Overall accuracy: 95.38\% \\
        Hemdan \textit{et al.}~\cite{hemdan:2020:covidxnet} & COVIDX-Net based on customized CNN & 50 chest X-rays (25 COVID-19 cases and 25 normal cases) & F1-score (Normal): 0.89, F1-score (COVID-19): 0.91 \\
        Narin \textit{et al.}~\cite{narin:2020:automatic} & ResNet-50/Inception-v3  & 100 chest X-rays across two classes (Normal: 50; COVID-19: 50) & Overall accuracy: 98.0 \% \\
        Ozturk \textit{et al.}~\cite{ozturk:2020:DarkNet} & DarkCovidNet & 1125 chest X-rays (125 COVID-19, 500 normal, and 500 pneumonia cases) & Overall accuracy for 3-class classification: 87.02\% \\
        Khan \textit{et al.}~\cite{khan:2020:CoroNet} & CoroNet & 3084 Chest X-rays from Kaggle (290 COVID-19, 1203 normal, 931 viral pneumonia and 660 bacterial pneumonia) & Accuracy (4-class): 89.6\%, Accuracy (3-class): 95.0\%. \\
        \textbf{This study} & CoVScreen & 3,160 Chest X-rays (Normal: 1,580, COVID-19: 1,580) & Accuracy: 97.46\%, AUC: 0.9963 \\        
         \bottomrule
    \end{tabular} 
\end{table*}
 
\section{Discussion}\label{discussion_section}
In this study, we developed an AI system named CoVScreen for the diagnosis of COVID-19 disease. Due to the nature of diseases, certain diseases occur more often compared to other diseases in the actual clinical practice, leading to class-imbalance problem. In order to mitigate the effect of class-imbalance, techniques such as \emph{random over-sampling} and \emph{random under-sampling} can be utilised to balance the number of samples for each class. In case, it is not possible to balance out the number of samples for each class, techniques such as \emph{class-weighting} scheme can be adopted in the objective function. Also, regularisation techniques such as \emph{dropout} can be incorporated in the network architecture. Data augmentation techniques such as rotation, scaling, flipping, shearing, and adding noise can also be applied during training in order to make model learn variation in the data and to get more number of under-represented class on the fly during training phase.

Although we try to mitigate model bias and overcoming shortcomings of various existing studies, we have not considered factors which should also have been taken care of while doing experimental design. Based on the experimental results and data analysis, it is clear that screening and diagnosis of a medical condition involves more than a mere classification task. Diagnosis of disease is a challenging problem compared to classifying an image as either a cat or dog, since there are subtle visual features for multiple medical concepts. Hence, it is important to use model interpretability and visualisation techniques, in addition to quantitative scores, to check which part of the chest X-ray model is paying attention to while making particular predictions. These visual explanations helps to align model predictions with what a radiologist would see in chest radiographs during radiographic study. In a study by \cite{Arias-Londono:2020:AI_applied_to_CXRs}, authors found that despite model's high performance in terms of accuracy score, the model is not paying attention to significant areas of diagnosis interest. Despite paying attention to significant areas of diagnostic interest in chest X-rays, the model is pointing towards corners of the images, sternum, and clavicles. They hypothesise that this may be because medical imaging metadata is embedded at the corners of the image. Because the model is learning the presence of metadata, the model is not classifying particular class based on disease radiographic features. This clearly indicates that the model is highly biased towards data characteristics rather than radiographic features of a medical condition. 

Another considerable factor which can have significant effect on the model performance is the projection of chest radiographs. Given that Posterior Anterior (PA) projections are often acquired in erect position, it is highly likely that chest X-rays having PA projections are captured to examine healthy or patients with mild symptoms of a particular medical condition. On the other hand, the Anterior Posterior (AP) projection are often acquired in supine position, it is likely that chest X-rays having AP projection are captured to examine severe cases of a disease or elderly patients who are confined to bed. The patients lying in bed are more likely to have medical support devices such as electrocardiogram electrodes and pacemakers, which can again bias model performance~\cite{Burlacu:2020:curbing_the_aI}. Thus, acquisition protocol can also affect the model performance. 

Based on our experiments and existing studies, it is found that model often do misclassify while distinguishing COVID-19 and viral pneumonia. Based on the hierarchical taxonomy of diseases, the \emph{SARS-CoV-2} (COVID-19) falls under the \emph{viral} diseases. Since pneumonia can also be caused due to viral infection, the radiographic features of \emph{viral pneumonia} and \emph{COVID-19 pneumonia} overlap, making model to possible misclassification. A study by ~\cite{Bai:2020:performance_of_radiologists} to validate the performance of seven radiologists in differentiating COVID-19 pneumonia from Non-COVID-19 pneumonia indicates that the sensitivities of three radiologists from China and four radiologists from the United States have range from $67\%$ to $97\%$ (mean sensitivity of $80.42\%$) and specificity have range from $7\%$ to $100\%$ (mean specificity of $83.71$). Authors found that compared to Non-COVID-19 pneumonia, the COVID-19 pneumonia is more likely to have a peripheral distribution ($80\%$ vs. 57\%), ground-glass opacity ($91\%$ vs. $68\%$), fine reticular opacity ($56\%$ vs. $22\%$), and vascular thickening ($59\%$ vs. $22\%$).

Since the performance of a diagnostic model for COVID-19 is highly dependent on the severity of COVID-19 infection, it is important to also report severity of disease in dataset for which results are reported. Based on various studies to analyse the temporal changes in radiographic features of the COVID-19 disease, it has been found that various radiographic features including \emph{ground-glass opacity}, \emph{reticular alterations}, \emph{peripheral infiltrates}, and \emph{consolidations} are the most common findings. Research studies~\cite{Wong:Lam:2020:frequency_and_distribution} also found that peak severity on CXRs is around $10$ to $12$ days from the onset of symptoms. However, once the disease reaches its peak in terms of severity, the consolidations reduce their density and course of the disease improves. This indicates that the COVID-19 disease has a particular evolution pattern, which can potentially be captured by automated algorithms to stratify COVID-19's stage in terms of the severity level. 

Although research towards automated diagnosis of COVID-19 using chest radiographs is in its infancy, the experimental results based on this and other studies shows the potential of learning algorithms in diagnosing COVID-19 disease. Various research directions, including understanding the mechanism of evolution of radiographic features in chest radiographs, temporal changes in radiographic features over the course of disease, automated methods to diagnose COVID-19 disease, and finding salient regions to point COVID-19 lesions using various interpretable and explainable tools, can be pursued. The overall goal of computer-aided diagnosis tools for COVID-19 can certainly expedite clinical workflow, augment radiologists' by providing them a ``second opinion", and reducing workload by automating tedious and laborious tasks such as quantitative analysis. Also, there is a need to have rigorous validation of these automated tools in real clinical settings so that it can bring insights about model's actual performance.  Finally, there is a need to have collaborative team involving radiologists, clinicians, and computer scientists, so that domain knowledge and trust of healthcare professional can be incorporated in such computer-aided diagnosis and screening tools.

\section{Limitations and future work}\label{limitations_and_future_work_section}
The goal of this study is to highlight the need of more thoughtful experimental design for the diagnosis of COVID-19 disease using chest radiographs. We highlighted various limitations in existing studies by showing results under different experimental settings. There are certain factors, which we have not included in our study, either due to lack of available metadata or lack of certain attributes in the available databases. First, no information is available about the difference in the period of time between onset of symptoms and acquisition of chest radiographs. This leads to bias in results given severe cases of COVID-19 report better performance compared to mild and moderate cases. Second, we have not performed a randomised and controlled study. Given this is a retrospective study, there is the possibility of bias. Third, demographic features such as age are not considered. Given that the performance of model may vary depending on the demographic characteristics of the target population, the model may exhibit higher performance on elderly people and people with impaired immune system, whereas model performance may be lower in young population. The curation of a large-scale dataset for COVID-19 diagnosis having linked chest X-rays with clinical information is highly sought after to validate diagnostic accuracy of automated methods. Fourth, our dataset is sourced from multiple locations across the globe and no study has been done towards transferability of AI models across different geographic locations; it is hard to determine whether the developed algorithm can be generalised to even broader population distributions over different regions and continents. Finally, unlike the RT-PCR test, which leads to a specific diagnosis of COVID-19, the chest radiographs findings are not specific enough to confirm COVID-19. Although medical imaging can point to signs and severity of infection, they can sometime ignore other causes of infection such as seasonal flu, which is very common at certain times of year.

We recommend the usage of any diagnostic model in conjunction to radiologist' findings based on chest radiographs for the diagnosis of COVID-19 disease. In conclusion, the combination of chest radiographs with the proposed CoVScreen deep learning algorithm has the potential as an accurate method to improve the accuracy and timeliness of the radiological interpretation of COVID-19 infection.

\section{Conclusions}
In this paper, we highlighted common pitfalls and recommendations for the diagnosis of COVID-19 using chest X-rays. We first provided limitations of existing studies in terms of curated dataset, data quality, and reported results. We then highlighted the need of thoughtful data curation and experimental design for the diagnosis of COVID-19 using AI tools. The proposed model is capable of providing accurate diagnostics for COVID-19 diagnosis given the importance of early detection of COVID-19 to avoid community transmission. AI tools for COVID-19 can augment radiologists or front line clinicians so that COVID-19 can be detected and monitored more quickly and objectively. Our research findings can contribute to the fight against diseases such as COVID-19 and help increase acceptance and adoption of AI-assisted applications in the clinical practice.
\section*{Sources of COVID-19 chest X-rays} 
Here, we provide various sources of original databases containing chest X-rays diagnosed with COVID-19 and as confirmed by the real-time reverse transcription polymerase chain reaction (RT-PCR) test. 

\begin{enumerate}
    \item IEEE8023/COVID-ChestXray-Dataset \\
    \url{https://github.com/ieee8023/covid-chestxray-dataset/}
    
    \item Figure 1 COVID-19 chest X-ray dataset initiative \\
    \url{https://github.com/agchung/Figure1-COVID-chestxray-dataset}
    
    \item COVID-19 chest imaging cases in a hospital in Spain \\
    \url{https://twitter.com/ChestImaging/status/1243935116541509632}
    
    \item SIRM COVID-19 Database \\
    \url{https://www.sirm.org/category/senza-categoria/covid-19/}
    
    \item Radiopedia \\
    \url{https://radiopaedia.org/articles/covid-19-4}
    
    \item COVIDGR-1.0 Dataset  \\
    \url{https://github.com/ari-dasci/OD-covidgr}
    
    \item Hannover Medical School COVID-19 Database \\
    \url{https://figshare.com/articles/COVID-19_Image_Repository/12275009/1}
    
    \item European Society of Radiology (ESR) COVID-19 cases \\
    \url{https://www.eurorad.org/advanced-search?search=COVID}
    
    \item Peer-reviewed publications \\
    \url{https://github.com/armiro/COVID-CXNet}
    
    \item BIMCV COVID19 Database  \\
    \url{https://bimcv.cipf.es/bimcv-projects/bimcv-covid19/}
    
    \item COVID-19 Chest X-ray image repository \\
    \url{https://github.com/armiro/COVID-CXNet}
\end{enumerate}

\begin{acknowledgements}
This research was undertaken with the assistance of resources from the National Computational Infrastructure (NCI) supported by the Australian Government.
\end{acknowledgements}

\noindent\textbf{Conflict of Interest}: There are no conflicts of interests in this work.

\bibliographystyle{spmpsci}      
\bibliography{references.bib}   
\end{document}